\documentclass[journal,romanappendices,twoside]{IEEEtran}

\usepackage{cite}
\usepackage{graphicx}
\graphicspath{{fig/}{img/}{fig/pdf/}}
\usepackage{amsfonts}
\usepackage{amssymb,amsmath}
\usepackage{dsfont}
\usepackage{booktabs}
\usepackage{calc}
\usepackage{ifthen}
\usepackage[dvipsnames,usenames]{color}
\usepackage{acronym}
\usepackage{nicefrac}
\usepackage{enumitem}
\usepackage{units}
\usepackage{soul}
\usepackage[absolute]{textpos}
\usepackage{comment}
\usepackage[demo]{pdfpages}

\newacro{TF}{Time-Frequency}
\newacro{FW}{Frequency Warping}
\newacro{TW}{Time Warping}
\newacro{SWF}{Sample Without Filtering}
\newacro{SAF}{Sample After Filtering}
\newacro{NUFFT}{Non Uniform Fast Fourier Transform}
\newacro{CQT}{Constant-Q Transform}

\newacro{SAS}{Synthetic Aperture Sonar}
\newacro{AUV}{Autonomous Underwater Vehicle}
\newacro{INS}{Inertial Navigation System}
\newacro{DPCA}{Displaced Phase Center Antenna}
\newacro{PCA}{Phase Center Approximation}
\newacro{EPCA}{Effective Phase Center Approximation}
\newacro{SIMO}{Single Input Multiple Output}
\newacro{SISO}{Single Input Single Output}
\newacro{SAR}{Synthetic Aperture Radar}
\newacro{NUFFT}{Non-Uniform Fast Fourier Transform}
\newacro{Tx}{Transmitter}
\newacro{Rx}{Receiver}
\newacro{PSF}{Point Spread Function}
\newacro{MIMO}{Multiple input Multiple Output}

\begin{document}

\title{A New Framework for\\Synthetic Aperture Sonar Micronavigation}

\author{Salvatore~Caporale and Yvan~Petillot
\thanks{S. Caporale  and Y. Petillot are with the Institute of Sensors, Signals and Systems, Heriot-Watt University, Edinburgh, Scotland, UK (e-mail: s.caporale@hw.ac.uk; y.r.petillot@hw.ac.uk).}}


\markboth%
{Caporale \MakeLowercase{\textit{et al.}}: A New Framework for Synthetic Aperture Sonar Micronavigation}
{Caporale \MakeLowercase{\textit{et al.}}: A New Framework for Synthetic Aperture Sonar Micronavigation}

%


\maketitle

\begin{abstract}
Synthetic aperture imaging systems achieve constant azimuth resolution by coherently summating the observations acquired along the aperture path. At this aim, their locations have to be known with subwavelength accuracy. In underwater \ac{SAS}, the nature of propagation and navigation in water makes the retrieval of this information challenging. Inertial sensors have to be employed in combination with signal processing techniques, which are usually referred to as micronavigation. In this paper we propose a novel micronavigation approach based on the minimization of an error function between two contiguous pings having some mutual information. This error is obtained by  comparing the vector space intersections between the pings orthogonal projectors. The effectiveness and generality of the proposed approach is demonstrated by means of simulations and by means of an experiment performed in a controlled environment.
\end{abstract}


\IEEEpeerreviewmaketitle
\acresetall

\section{Introduction}
\label{sec:intro}

\ac{SAS} systems share with \ac{SAR} many practical and theoretical aspects, as they were originally introduced by moving the synthetic aperture paradigm from radar to sonar. Therefore, most image formation algorithms which have been conceived for \ac{SAR} have been also considered in \ac{SAS} literature\cite{CafforioPratiRocca1991}. Despite the overlapping concepts such as range migration and range invariant resolution, underwater \ac{SAS} systems are operated in a much more challenging environment than \ac{SAR} as (i) the navigation in water is affected by non-negligible errors and cannot always rely on an external accurate positioning system, (ii) motion errors are comparable to the wavelength hence their effect on image formation algorithms is remarkably destructive (iii) due to relatively small sound velocity in water with respect to the range of interest, the desired along-track resolution cannot be achieved by means of an \ac{AUV} provided with a single \ac{Tx} and \ac{Rx} and moving at a reasonable along-track speed~\cite{BillonPinto1995,HayesGough2009}.

For the above mentioned reasons, \ac{SAS} systems are usually equipped with an accurate \ac{INS} to recover the real navigated trajectory. Moreover, each ping consists of a single \ac{Tx} and an array of \acp{Rx}, allowing for a higher along-track sampling rate, hence a higher along-track speed~\cite{BellettiniPinto2009}. However, from ping to ping a certain degree of redundancy is imposed to perform data based motion estimation known as \ac{DPCA} which might be coupled with the \ac{INS} for a more accurate motion compensation~\cite{BellettiniPinto2002,CallowHayesGough2009}.
A typical \ac{SAS} system and the way it operates are illustrated in Fig.~\ref{fig:ssy}. Other approaches based on autofocus techniques have been also presented ni the literature~\cite{gough2004,marston2015,herter2016}.

In a stripmap imaging system where the \ac{AUV} moves along a straight trajectory at constant speed, major issues are caused by cross-range displacements. In fact, the sensitivity of the \ac{SAS} \ac{PSF} is remarkably higher along the cross-track than along the along-track~\cite{HayesGough2009}. The cross-track motion has to be known with subwavelength accuracy, whereas the along-track motion has to be in the order of portion of the sampling step. Hence, micronavigation algorithms focus on finding cross-track errors whereas it is generally assumed that the along-track locations are either correct or can be estimated by the navigation system~\cite{BellettiniPinto2009}. 
The theoretical foundation of several currently employed approaches relies on (i) \ac{PCA}, (ii) the hypothesis that the \ac{AUV} rotation angle is \emph{small} such that it can be characterized as a linearly increasing delay along the array elements and (iii) contiguous pings have some superimposed phase centers, such that the mutual delays between the corresponding tracks can be estimated by finding their correlation peak. Many challenges in \ac{SAS} micronavigation still have to be addressed~\cite{HayesGough2009}. Those include (i) dealing with non straight trajectories, such as circular ones~\cite{marston2012}, (ii) reducing the hardware cost by relying on less accurate navigation and (iii) analysing theoretical limits in performing unsupervised motion compensation.

Following the work in~\cite{caporale2016}, we here exploit the superimposition of some phase centers as for \ac{DPCA}. However, we assume that they can be arbitrarily interlaced rather than exactly superimposed.
Then, the vector space intersection between the space corresponding to the two pings is described as a functional of the hypothetical displacement. Each ping is employed to compute a pair of displacement-dependent outputs being the projections on the intersection subspace. A proper convex error function between those outputs features a minimum in correspondence of the ping-to-ping displacement, so that it can be identified by means of an optimization. This approach  is computationally demanding due to the evaluation of an error function involving the recomputation of the projecting operators. However, it requires less a priori knowledge and is remarkably less restrictive in comparison to \ac{DPCA}.


The paper is organized as follows. Section \ref{sec:acmod} reviews the acoustic model and \ac{SAS}, while Section \ref{sec:motcomp} illustrates the proposed motion compensation procedure. Results from synthetic data and from a real experiment are shown in Section \ref{sec:sim} and \ref{sec:exp} respectively and some conclusions are drawn in Section \ref{sec:concl}.

\section{Acoustic and SAS Model}
\label{sec:acmod}

In this Section we first review the observation model for a single \ac{Tx} and \ac{Rx} acoustic system when a generic signal is given as input, then we show how this can be employed to retrieve the scene reflectivity when multiple observations are combined together. A simplified 2-D model is considered by means of the assumption that (i) the seabed reflectivity can be represented as a function on a plane and (ii) the 2-D scene reflectivity can be obtained as the projection of the 2-D seabed reflectivity on a slanted plane. Moreover, all the sensors are assumed to be still. Finally, \ac{SAS} is described as a collection of such single \ac{Tx} and \ac{Rx} systems.

\subsection{Acoustic Model}

\begin{figure}[t]
	\centering%
	\includegraphics[width=\columnwidth]{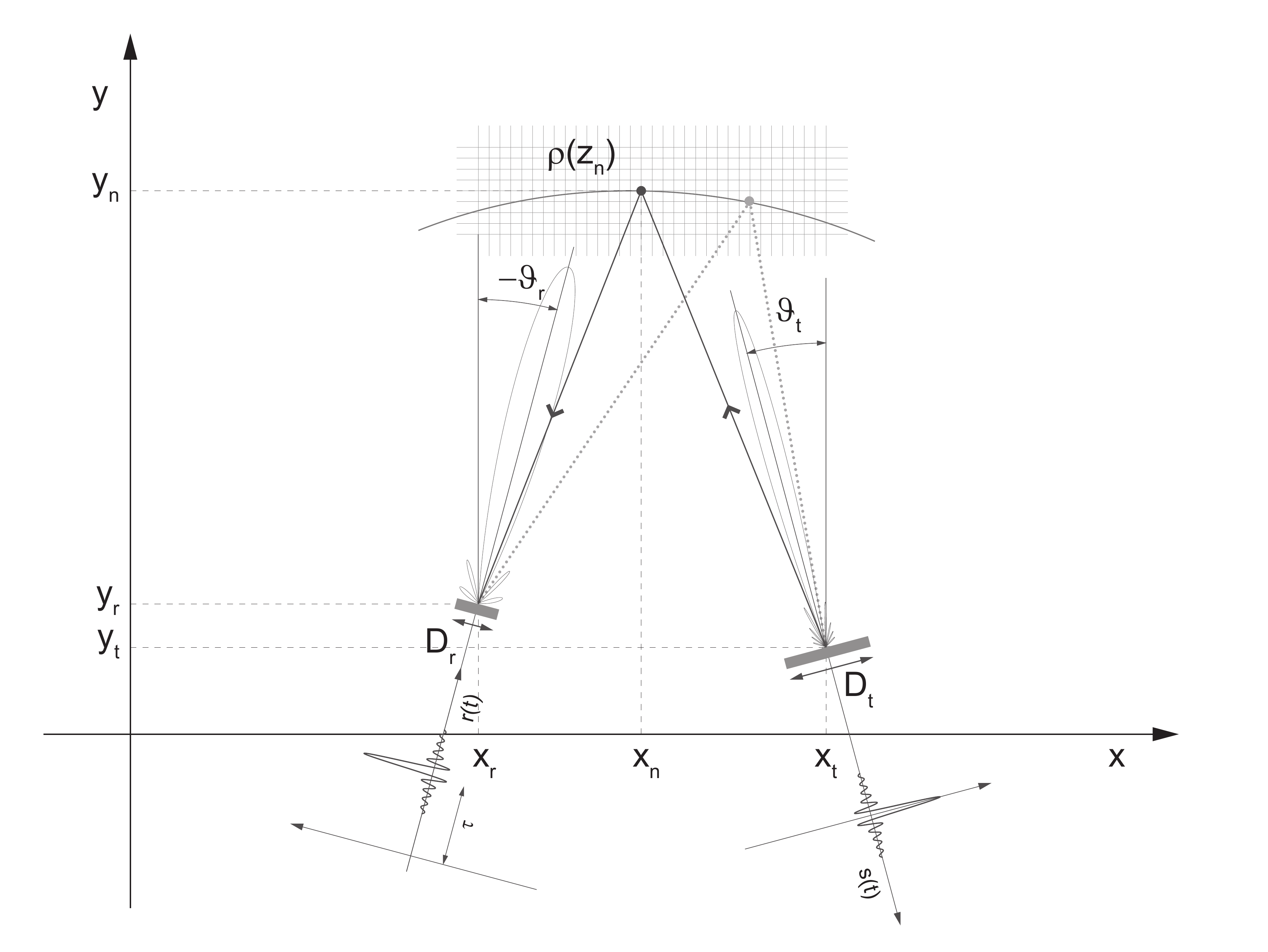}%
	\caption{%
		The adopted single \ac{Tx} and \ac{Rx} bistatic 2-D acoustic model. The figure also shows two different propagation paths corresponding to two reflectors sharing the same round trip time, thus lying on the ellipsis having the \ac{Tx} and the \ac{Rx} as foci.
	}%
	\label{fig:siso}
\end{figure}

Given a \ac{Tx} at position $z_t=(x_t,y_t)$ of length $D_t$ and orientation $\vartheta_t$ and a hydrophone at position $z_r=(x_r,y_r)$ of length $D_r$ and orientation $\vartheta_r$, the system response to an input complex passband function $s(t)$ with respect to a scene whose complex reflectivity is expressed by $\rho(z)$, with $z=(x,y)$ can be obtained by:
\begin{equation}\label{eq:acmod}
r(t)=\int_{z}\rho(z)\,\alpha(z_t,z_r,z,\vartheta_t,\vartheta_r)\,s(t-\tau(z_t,z_r,z))\,\mathrm{d}z
\end{equation}
where $\tau$ is the delay corresponding to the propagation from $z_t$ to $z$ and from $z$ to $z_r$, $\alpha$ is an attenuation factor taking into account (i) the attenuation due to the propagation distance  from $z_t$ to $z$ and from $z$ to $z_r$ and (ii) the transmission and reception radiation patterns. With regard to the attenuation, we consider the exploding sources model where the $\alpha^2\propto\delta(z_t,z)+\delta(z_r,z)\propto\tau(z_t,z_r,z)$ rather than $\alpha^2\propto{\delta(z_t,z)\delta(z_r,z)}$, where $\delta$ is the distance operator. The model \eqref{eq:acop} is represented in Fig.~\ref{fig:siso}.

When $s(t)$ is a narrowband impulse modulated at frequency $f_0$, the above model can be used for observing $\rho(z)$. A suitable sampling grid $z_n=(x_n,y_n)$ matching the available bandwidth is introduced so that the reflectivity is represented as the input vector $\rho(z_n)$. By setting $G(z_n)$ as the Green's function of the system
\begin{equation}
G(z_n)=\alpha(z_t,z_r,z_n,\vartheta_t,\vartheta_r)e^{-j2\pi{}f_0\tau(z_t,z_r,z_n)},
\end{equation}
%
%
the following output time-sampled signal $\phi(t_m)$ can be obtained as
\begin{equation}\label{eq:acop}
\phi(t_m)=A(t_m,z_n)\,G(z_n)\rho(z_n)
\end{equation}
where the matrix $A(t_n,z_n)$ performs the discrete-space integration of equation \eqref{eq:acmod}, which is basically done by an interpolation kernel and a summation. From a numerical point of view, $A(t_m,z_n)$ is usually defined by its transpose one, since defining an interpolation along the intrinsic one dimensional time axis $t_m$ is easier than doing it over the fictitious space axis $z_n$. We consider a \ac{NUFFT} based time interpolation \cite{FesslerSutton2003,caporale2007}, having an approximately linear complexity and high accuracy.

The integral in \eqref{eq:acmod} and its space discrete equivalent in \eqref{eq:acop} clearly perform a dimensionality reduction from the two-dimensional space $z_n$ to the one dimensional space $t_m$.  In fact, all points lying on the ellipsis having $z_t$ and $z_r$ as foci are intrinsically ambiguous with respect to the \ac{Tx} \ac{Rx} pair. This is also represented in Fig.~\ref{fig:siso} where we highlighted a propagation path (dotted line) having the same delay as the one relative to the generic point $z_n$. As a consequence, inverting \eqref{eq:acop} recovers the reflectivity $\rho(z)$ affected by defocusing along ellipsis. 
Due to the radiation patterns, this defocusing increases with range.

\subsection{SAS model}

The working principle of a synthetic aperture consists in performing multiple observations at prescribed \ac{Tx} and \ac{Rx} locations along a (usually straight) path, called along-track or cross range. By doing so, a longer aperture is emulated and a narrower beam is obtained in the direction orthogonal to the track, called range. The emulation is performed by signal processing and allows for getting a constant resolution along range.

Given a set of \acp{Tx} and \acp{Rx} $z_{l,t}$ and $z_{l,r}$, $l\in\mathds{Z}$, the observation model of \ac{SAS} can be written as the column vector of $\phi_l(t_m)=A_l(t_m,z_n)\,G_l(z_n)\,\rho(z_n)$.
%
%
In case each observation is a monostatic system, i.e. $z_{l,t}=z_{l,r}=z_{l}$, and $z_{l}$ are uniformly spaced along the cross range direction by $D/4$, with $D=\max({D_t},{D_r})$, then $\rho(z_n)$ can be approximately recovered with a cross range resolution equal to $D/2$ by the transpose observation operator, also know as backprojection
\begin{equation}\label{eq:bckprj}
	\rho(z_n)\approx\sum_{l\in\mathds{Z}}G^*_l(z_n)A_l^\dag(z_n,t_m)\phi_l(t_m)
\end{equation}
which has been conveniently rewritten as a sum of backprojections of each single \ac{Tx} and \ac{Rx} system.
As for getting a cross range resolution equal to $D/2$ a sampling on the along-track by $D/4$ has been considered, the oversampling ratio is equal to 2. Many variations of this setup have been considered in the literature in order to decrease the oversampling ratio and employ different reconstruction techniques~\cite{BellettiniPinto2002}.


Practical \ac{SAS} systems are mounted on an \ac{AUV}. Hence, the speed $v$ of the vehicle has to be set according to the wanted along-track sampling and the desired maximum range. For instance, in case of an \ac{AUV} equipped with a single \ac{Tx} and \ac{Rx}, $v\leq{}\nicefrac{D}{4\max(\tau(z_{l,t},z_{l,r},z_n))}=\nicefrac{Dc}{8R}$, where $R$ is the maximum range. For $R=\unit[150]{m}$ and $D=\unit[5]{cm}$, we get $v=\unit[6.25]{cm/sec}$. A feasible speed is obtained by using an array of $N$ equispaced \acp{Rx} at distance $L=D/2$ and a single \ac{Tx}. By doing so, a single transmitted ping allows for $N$ observations, thus gaining a $N$ factor on the speed. Such a \ac{SAS} system is a collection of bistatic single \ac{Tx} and \ac{Rx} systems. If the range of interest is sufficiently large with respect to the wavelength, it can be approximately modelled as a collection of monostatic systems such that $\bar{z}_l=({z}_{l,t}+{z}_{l,r})/2$. This model is referred to as \ac{PCA}.

\section{Motion Estimation}
\label{sec:motcomp}

\begin{figure}[t]
	\centering%
	\includegraphics[scale=1.068,trim=0 10pt 0 10pt]{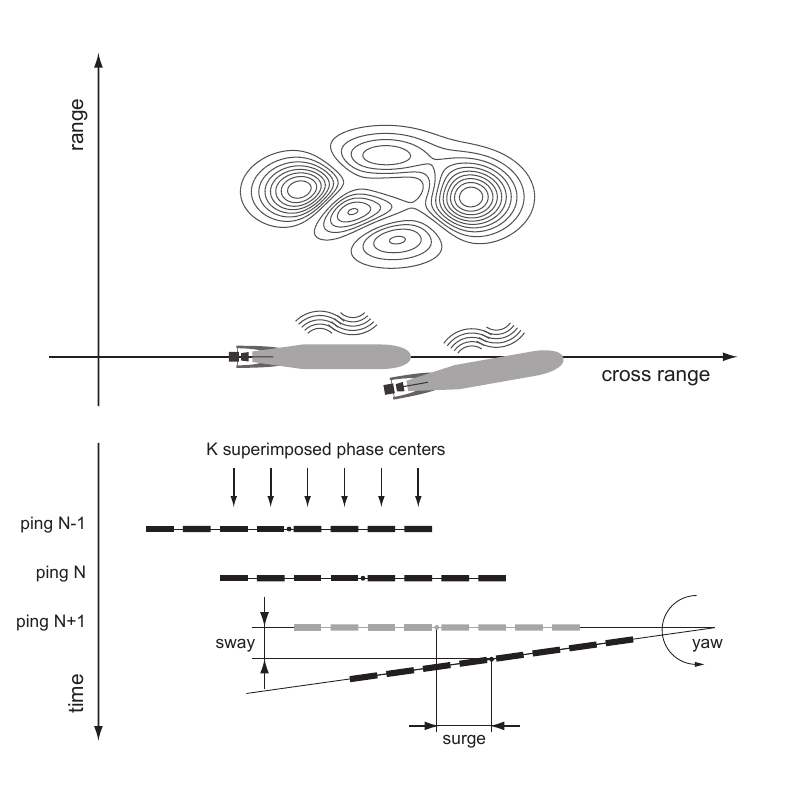}%
	\caption{%
		A typical \ac{SAS} setup design in order to cope with motion errors. At each ping the vehicle should be located in prescribed positions, as it is happening for ping $N-1$ and $N$. Ping $N+1$ features displacements along both range and cross-range directions together with rotation with respect to moving path. The overlap between subsequent pings gives the possibility of retrieving the differential motion errors. Unlike other approaches requiring no along-track error and small yaw, no restrictions are assumed here.
	}%
	\label{fig:ssy}
\end{figure}

The motions of an \ac{AUV} in a 3-D space are represented by means of 3 linear motion parameters being heave, sway and surge, and 3 rotation motion parameters being pitch, roll and yaw. As the acoustic model has been described in 2-D, only the effective projections of those on the slanted plane are needed, i.e. surge and sway as effective linear motions and yaw and roll as effective rotations. Roll only affects the radiation pattern over range and can be neglected. Without loss of generality, we here refer to surge as the motion error with respect to the prescribed linear trajectory at constant speed. Surge, sway and yaw are illustrated in Fig.~\ref{fig:ssy}.

The \ac{SAS} image formation requires an accuracy by approximately $\nicefrac{\lambda}{10}$ on $\bar{z}_l$, where $\lambda$ is the wavelength. Underwater navigation is not accurate enough to allow this accuracy for the employed frequency band, usually $\unit[100-300]{kHz}$. As a consequence, a motion estimation has to be performed to guarantee that the acquired observations are coherently combined. This is usually achieved by combining navigation information coming from physical inertial sensors place on the \ac{AUV}, that is the \ac{INS}, with information extracted from the acquired data by introducing redundancy from ping to ping. Instead of moving the \ac{AUV} at the maximum allowed speed, a reduced speed is adopted in order to have a certain number of $K<N$ superimposed phase centers, as represented in Fig.~\ref{fig:ssy}. This approach is usually referred to as \ac{DPCA}. Under the assumptions that (i) \ac{PCA} holds, (ii) surge is negligible and (iii) yaw is \emph{small}, the tracks from two contiguous pings referred to the same phase center differ for a time delay and the delays of the $K-N$ superimposed phase centers are an affine function of the sway and yaw. The delays can be estimated by correlating the corresponding tracks, thus also sway and yaw can be obtained. Interested readers are referred to~\cite{BellettiniPinto2002} for more details. An approach to tackle surge estimation is presented in~\cite{hunter2016}.

We here propose a technique loosening the three assumptions above, more specifically (j) \ac{PCA} is replaced with a less restrictive approximation, (jj) the along-track motion is such that an unknown non-integer number of superimposed phase centers are present and (jjj) yaw is only small enough to guarantee that the area illuminated by two contiguous pings has a non-null intersection.

\subsection{Ping to Ping Displacement Estimation}
\label{sec:pingping}

Given that the receiving array is composed by $N$ equispaced \acp{Rx}, the raw data obtained by collecting the reflections from the scene of a single impulse will be referred to as \emph{ping}. For the sake of the stripmap \ac{SAS} under consideration, the \ac{AUV} is moving along a straight trajectory affected by motion errors. The scene is observed by means of the start-and-stop approximation. Given the trajectory, each ping can be represented as a collection of bistatic systems identified by the array midpoint and orientation and the \ac{Tx} position and orientation:
\begin{eqnarray*}
	\sigma_r^{(p)}&=&(x_a^{(p)},y_a^{(p)},\vartheta_r^{(p)})\\
	\sigma_t^{(p)}&=&(x_t^{(p)},y_t^{(p)},\vartheta_{t}^{(p)}).
\end{eqnarray*}
where $x_a$ and $y_a$ are the coordinates of the midpoint of the receiving array elements. We here assume that the relative position of $\sigma_t^{(p)}$ with respect to $\sigma_r^{(p)}$ is known. In a real scenario, the displacement between the transmitter and the array midpoint is time-dependent as the \ac{AUV} moves between the transmission instant and the reception relative to the considered range interval (see~\ref{sec:txrx}). 
We also introduce the operator $T^{(p)}$ representing the observation model relative to ping $p$, so that $T^{(p)}\rho$ gives the raw data relative to ping $p$. We also set
%
\begin{equation}\label{eq:pseudoinv}
	\breve{\rho}^{(p)}=\tilde{T}^{(p)-1}\,T^{(p)}\,\rho
\end{equation}
where $\breve{\rho}^{(p)}$ represents the recovered reflectivity at ping $p$ by means of the pseudoinverse operator $\tilde{T}^{(p)-1}$, where $\tilde{T}$ is the available computable approximation of the real observation operator $T^{(p)}$. Mathematically, $\tilde{T}^{(p)}$ is the column matrix of the real observation matrices $A_{l,\gamma}G_{l,\gamma}$, $l=1,\ldots,N$ (see \eqref{eq:bckprj}).
Then, we define the orthogonal projector on the subspace identified by ping $p$
\begin{equation*}
	Q^{(p)}=(\tilde{T}^{(p)})^{-1}\tilde{T}^{(p)}.
\end{equation*}
Given another generic ping $q$ and its orthogonal projector $Q^{(q)}$, the projection on the intersection of the subspaces identified by $Q^{(p)}$ and $Q^{(q)}$ can be obtained by
\begin{equation}\label{eq:intersect}
	\psi^{(p,q)}=\psi^{(q,p)}=\lim_{i\to\infty}(Q^{(q)}Q^{(p)})^i\rho=\lim_{i\to\infty}(Q^{(p)}Q^{(q)})^i\rho.
\end{equation}
By taking advantage of \eqref{eq:pseudoinv}, an approximation of $\psi^{(p,q)}$ can be obtained by starting either from $\breve{\rho}^{(p)}$ or $\breve{\rho}^{(q)}$. In fact
\begin{eqnarray}
	\breve{\rho}^{(p)}&\simeq&\tilde{T}^{(p)-1}\tilde{T}^{(p)}\rho=Q^{(p)}\rho
	\label{eq:pseudoapproxp}
	\\
	\breve{\rho}^{(q)}&\simeq&\tilde{T}^{(q)-1}\tilde{T}^{(q)}\rho=Q^{(q)}\rho
	\label{eq:pseudoapproxq}
\end{eqnarray}
hence by highlighting the expressions \eqref{eq:pseudoapproxp}-\eqref{eq:pseudoapproxq} in \eqref{eq:intersect} we get
\begin{eqnarray}
	\psi^{(p,q)}&=&\lim_{i\to\infty}(Q^{(p)}Q^{(q)})^iQ^{(p)}\rho
	\nonumber
	\\
	&\simeq&\lim_{i\to\infty}(Q^{(p)}Q^{(q)})^i\breve{\rho}^{(p)}=\psi^{(p)}
	\label{eq:intersectpq}
	\\
	\psi^{(p,q)}&=&\lim_{i\to\infty}(Q^{(q)}Q^{(p)})^iQ^{(q)}\rho
	\nonumber
	\\
	&\simeq&\lim_{i\to\infty}(Q^{(q)}Q^{(p)})^i\breve{\rho}^{(q)}=\psi^{(q)}
	\label{eq:intersectqp}
\end{eqnarray}
and finally we get the relationship
\begin{equation}\label{eq:intersecteq}
	\psi^{(p)}\simeq\psi^{(q)}.
\end{equation}
The meaning of \eqref{eq:intersecteq} is that if the intersection operator is known, it is possible to identify the same \emph{intersection} image starting from the raw data obtained by two different pings. In case the operator $T^{(p)}$ and $T^{(q)}$ are orthogonal, both $\psi^{(p)}$ and $\psi^{(q)}$ are null, hence their computation is not relevant. Conversely, in case they are not orthogonal, the fact that $\|\psi^{(p)}-\psi^{(q)}\|\simeq{}0$, where $\|\cdot\|$ is a generic norm, can be exploited as a property to identify the intersection operator.

In order to apply the principle expressed in equation \eqref{eq:intersecteq}, we introduce operator $Q^{(p)}_0$ being a shifted/rotated version of $Q^{(p)}$ such that the center of mass of the \ac{PCA} is the origin of the axes. In more detail, by setting
\begin{eqnarray*}
	\bar{x}^{(p)}&=&(x_a^{(p)}+x_t^{(p)})/2\\
	\bar{y}^{(p)}&=&(y_a^{(p)}+x_t^{(p)})/2
\end{eqnarray*}
we get
\begin{eqnarray*}
	\sigma_{r,0}^{(p)}&=&(x_a^{(p)}-\bar{x}^{(p)},y_a^{(p)}-\bar{y}^{(p)},0)\\
	\sigma_{t,0}^{(p)}&=&(x_t^{(p)}-\bar{x}^{(p)},y_t^{(p)}-\bar{y}^{(p)},\vartheta_{t}^{(p)}-\vartheta_{r}^{(p)})
\end{eqnarray*}
and we set
\begin{equation*}
	\bar{\sigma}^{(p)}=(\bar{x}^{(p)},\bar{y}^{(p)},\vartheta_r^{(p)}).
\end{equation*}
By introducing the shift and rotation operator $S_\sigma$, where $\sigma$ specifies a triplet $(x,y,\vartheta)$, we finally get
\begin{equation*}
	Q^{(p)}=S_{\bar{\sigma}^{(p)}}Q^{(p)}_0S_{-\bar{\sigma}^{(p)}}.
\end{equation*}
Ping $q$ can be specified by $Q_0^{(q)}$ and $\bar{\sigma}^{(q)}$, or, alternatively, by $Q_0^{(q)}$ and $\bar{\sigma}^{(p)}$ and the differential displacement between $\bar{\sigma}^{(p)}$ and $\bar{\sigma}^{(q)}$, which will be referred to as $\bar{\sigma}^{(q,p)}$
\begin{eqnarray*}
	Q^{(q)}
	&=&
	S_{\bar{\sigma}^{(q)}}Q^{(q)}_0S_{-\bar{\sigma}^{(q)}}\\
	&=&
	S_{\bar{\sigma}^{(p)}}S_{\bar{\sigma}^{(q,p)}}Q^{(q)}_0S_{-\bar{\sigma}^{(q,p)}}S_{-\bar{\sigma}^{(p)}}
\end{eqnarray*}
with
\begin{eqnarray*}
	\bar{\sigma}^{(q,p)}&=&(\bar{x}^{(q)}-\bar{x}^{(p)},\bar{y}^{(q)}-\bar{y}^{(p)},\vartheta_r^{(q)}-\vartheta_r^{(p)})
	\\
	&=&(x^{(q,p)},y^{(q,p)},\vartheta^{(q,p)})
\end{eqnarray*}
hence the following is verified
\begin{equation*}
	\bar{\sigma}^{(p,q)}=-\bar{\sigma}^{(q,p)}.
\end{equation*}

It is worth noting that the differential displacement which has been highlighted is equal to the one which would be obtained by replacing the bistatic situation with its \ac{PCA}, nevertheless, this model maintains the capability of representing a bistatic system where the transmitter and the array have different orientations.

So, the vector $\bar{\sigma}^{(q,p)}$ fully identifies the displacement between ping $p$ and $q$ and represents the goal of the motion estimation procedure. In order to estimate it, we rewrite equations \eqref{eq:intersectpq} and \eqref{eq:intersectqp} by taking into account that $\breve{\rho}^{(p)}$ and $\breve{\rho}^{(q)}$ are not available, whereas their rotated ans shifted version by $S_{-\bar{\sigma}^{(p)}}$ can be computed
\begin{eqnarray*}
	S_{-\bar{\sigma}^{(p)}}\psi^{(p)}&=&\lim_{i\to\infty}(Q_0^{(p)}S_{\bar{\sigma}^{(q,p)}}Q^{(q)}_0S_{\bar{\sigma}^{(p,q)}})^{i}S_{-\bar{\sigma}^{(p)}}\breve{\rho}^{(p)}
	\\
	S_{-\bar{\sigma}^{(p)}}\psi^{(q)}&=&\lim_{i\to\infty}(S_{\bar{\sigma}^{(q,p)}}Q^{(q)}_0S_{\bar{\sigma}^{(p,q)}}Q_0^{(p)})^{i}S_{-\bar{\sigma}^{(p)}}\breve{\rho}^{(q)}.
\end{eqnarray*}
%
Since the displacement between $p$ and $q$ is unknown, the above equations can be rewritten with respect to a hypothetical displacement $\sigma=(x,y,z)$
\begin{eqnarray*}
	S_{-\bar{\sigma}^{(p)}}{\psi}^{(p)}_{{\sigma}}&=&\lim_{i\to\infty}(Q_0^{(p)}S_{{\sigma}}Q^{(q)}_0S_{-{\sigma}})^{i}S_{-\bar{\sigma}^{(p)}}\breve{\rho}^{(p)}
	\\
	S_{-\bar{\sigma}^{(p)}}{\psi}^{(q)}_{{\sigma}}&=&\lim_{i\to\infty}(S_{{\sigma}}Q^{(q)}_0S_{-{\sigma}}Q_0^{(p)})^{i}S_{-\bar{\sigma}^{(p)}}\breve{\rho}^{(q)}
\end{eqnarray*}
In general, for non-null $\breve{\rho}^{(p)}$ and $\breve{\rho}^{(q)}$ and non-null intersection operator, we have
\begin{equation*}
	{\psi}^{(p)}_{{\sigma}}={\psi}^{(q)}_{{\sigma}}
	\quad
	\Leftrightarrow
	\quad
	{\sigma}=\bar{\sigma}^{(q,p)}.
\end{equation*}
%
In practice, they will not be equal because of approximation errors. However,  ${\psi}^{(p)}_{{\sigma}}$ and ${\psi}^{(q)}_{{\sigma}}$ can be employed for building an error function. Since they represent wavefield reconstructions, they feature periodic-wise variations with period equal to half the wavelength. An error function based on those would also feature such oscillations, hence it is expected to be convex on a small hyperinterval around the displacement triplet $(x^{(q,p)},y^{(q,p)},\vartheta^{(q,p)})$:
\begin{equation}\label{eq:errph}
	\eta^{(q,p)}(x,y,\vartheta)=
	\big\|\,{\psi}^{(p)}_{{\sigma}}-{\psi}^{(q)}_{{\sigma}}\,\big\|_2.
\end{equation}
Conversely, their absolute values represents reflectivity amplitudes so an error function based on them is expected to be convex on a larger hyperinterval although it does not necessarily have its minimum in $(x^{(q,p)},y^{(q,p)},\vartheta^{(q,p)})$
\begin{equation}\label{eq:errmd}
	\zeta^{(q,p)}(x,y,\vartheta)=
	\big\|\,\big|{\psi}^{(p)}_{{\sigma}}\big|-\big|{\psi}^{(q)}_{{\sigma}}\big|\,\big\|_2.
\end{equation}
Functions $\zeta$ and $\eta$ will be referred to as \emph{modulus} and \emph{phase} error function respectively. The interval of convexity of the above functions will be discussed later in the paper. Normalized version of the above functions are more conveniently employed, where the normalization can be done with respect to either ${\psi}^{(p)}_{{\sigma}}$, ${\psi}^{(q)}_{{\sigma}}$ or the maximum between the two.

Hence, the estimation procedure can be performed by:
\begin{enumerate}[leftmargin=14pt]
	\item initialize $\sigma$: $\sigma\leftarrow(0,0,0)$
	\item minimize $\zeta^{(p,q)}$ starting from $\sigma$: $\sigma\leftarrow\min\zeta^{(p,q)}$
	\item minimize $\eta^{(p,q)}$ starting from $\sigma$: $\sigma\leftarrow\min\eta^{(p,q)}$
\end{enumerate}
The characteristics of $\zeta$ and $\eta$ cannot be analytically identified. A practical assessment of this proposed procedure will be provided in Section \ref{sec:sim}. The following considerations hold:
\begin{itemize}[leftmargin=14pt]
	\item the superimposition between phase centers is not required as the along-track displacement is also part of the search;
	\item the estimation relies on global similarities in the image domain rather than on track-by-track correlations, thus more robust;
	\item rotation estimation relies only on non-null vector space intersection between pings, hence \emph{large} rotation can be estimated.
\end{itemize}
However, there are some major computational burdens:
\begin{itemize}[leftmargin=14pt]
	\item each optimization iteration requires the update of the observation model of one of the two pings;
	\item both the orthogonal projector and the vector space intersection requires infinite iterations which must be properly truncated.
\end{itemize}
To decrease the computational load, the range of interest can be conveniently limited.

\subsection{Approximate Bistatic Model}
\label{sec:txrx}


%

\begin{figure}[t]
	\centering%
	\includegraphics[scale=.75,trim=0 10pt 0 10pt]{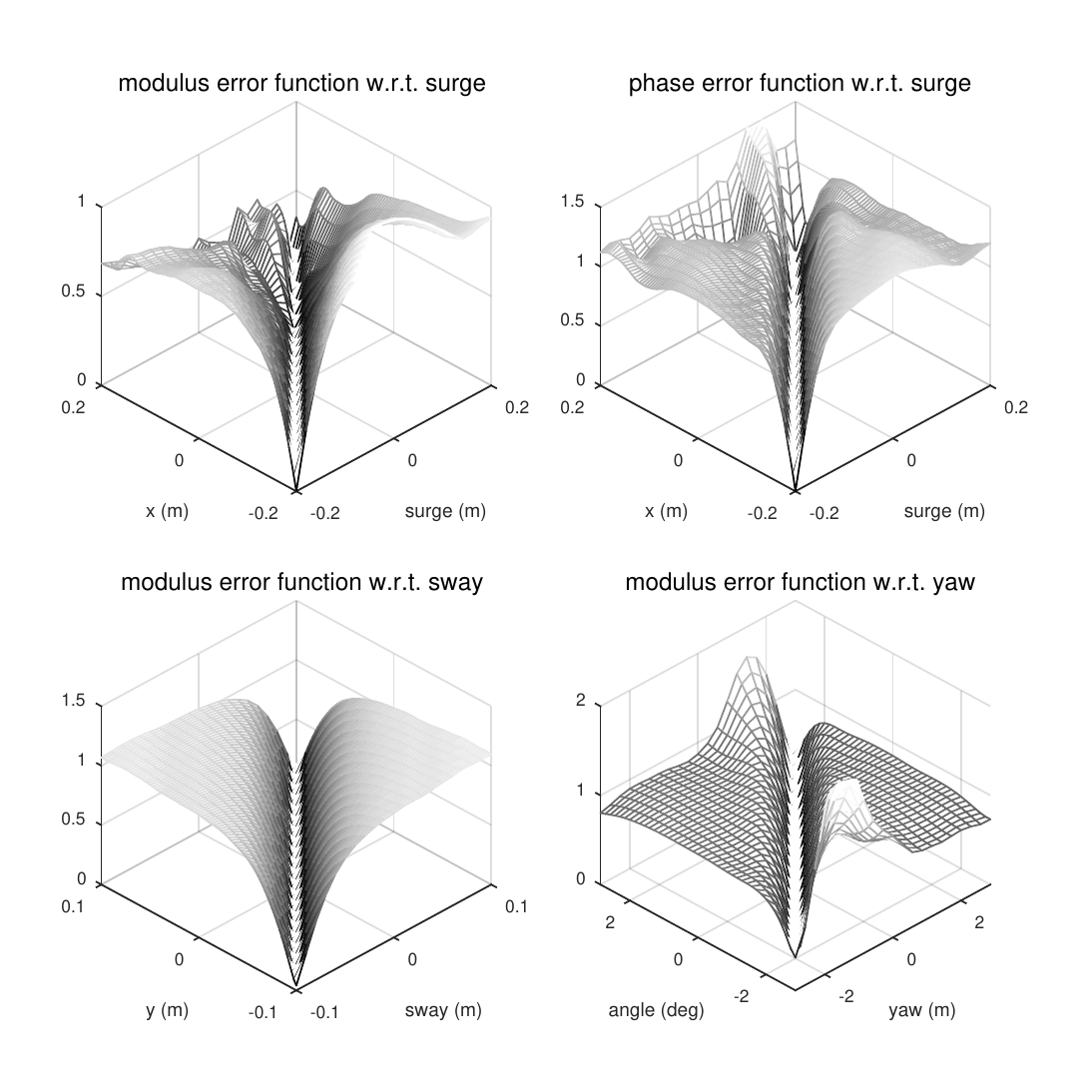}%
	\caption{%
		Error functions employed to estimate motion errors when only one component is considered at a time. For the surge, both the error based on projection modulus and the one considering also the phase are illustrated (top), where the first exhibits a wider convex interval around the minima, whereas the latter is more accurate. With respect to sway, the error features a smooth convex behaviour (bottom-left), while for yaw some local minima might occur which might hinder the convergence to the global minimum.
	}%
	\label{fig:srf}
\end{figure}

\begin{figure}[t]
	\centering%
	\includegraphics[scale=.75,trim=0 10pt 0 10pt]{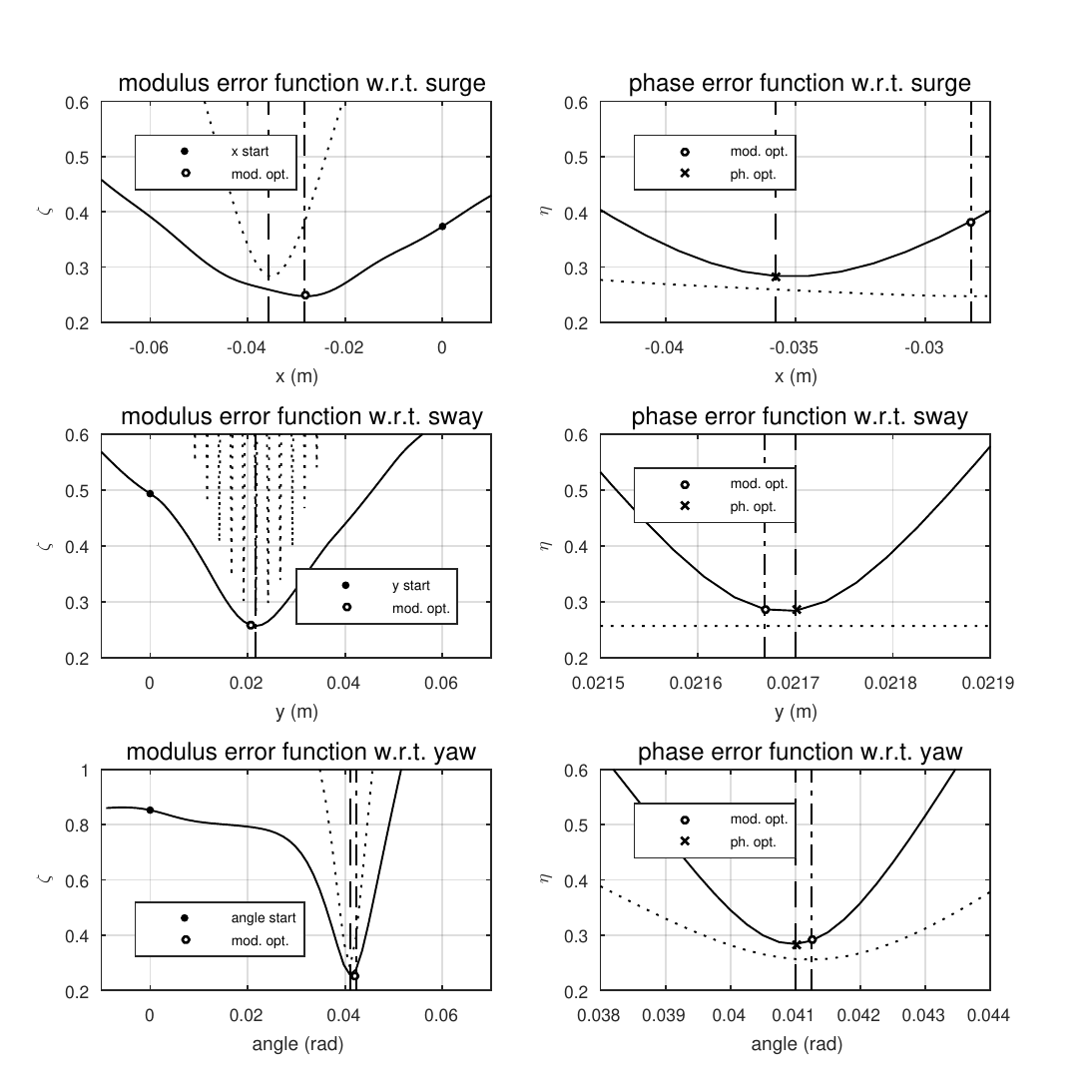}%
	\caption{%
		Error functions and estimation procedure for the proposed approach exemplified on surge, sway and yaw separately (top, middle and bottom respectively). First, an optimization is performed on the function $\zeta$ (left, solid line) and then the found minimum is refined by a further optimization on function $\eta$ (right, solid line). $\zeta$ allows for larger intervals of convexity, whereas $\eta$ leads to a higher accuracy.
		Dashed vertical lines identify the displacement to be estimated. Dash-dot lines identify the output of the first optimization.
	}%
	\label{fig:cuts}
\end{figure}

In the previous subsection we assumed each ping to be obtained as the output of a collection of $N$ bistatic systems under the start-and-stop approximation. As the estimation approach do not rely on the \ac{PCA}, we here introduce a more general approximation allowing for representing (i) the orientation of the \ac{Tx} with respect to the orientation of the array and (ii) the effective displacement along the cross-track between the \ac{Tx} and the array which is happening because the \ac{AUV} motion. The employment of bistatic systems has been shown to carry some benefits~\cite{lepage2002}, hence the capability of including them is beneficial.

With the assumption that the transmitted signal is ideally instantaneous (no pulse compression) and that the observed range interval tends to zero, $\sigma_r^{(p)}$ and $\sigma_t^{(p)}$ can be determined by sampling the effective \ac{AUV} trajectory on the slanted plane. According to \ac{PCA} are approximated by their average. Infinite different bistatic systems sharing the same \ac{PCA} can be identified. Among these, we approximate $\sigma_r^{(p)}$ and $\sigma_t^{(p)}$ by $\tilde{\sigma}_r^{(p)}$ and $\tilde{\sigma}_t^{(p)}$, such that
%
\begin{eqnarray}
	\tilde{\sigma}_r^{(p)}&=&\bar{\sigma}^{(p)}+\varsigma_r
	\label{eq:epcar}
	\\
	\tilde{\sigma}_t^{(p)}
	&=&\bar{\sigma}^{(p)}+\varsigma_t+(0,0,\vartheta^{(p)}_t-\vartheta^{(p)}_r)
	\label{eq:epcat}
\end{eqnarray}
where $\varsigma_r=(x_a,y_a,0)$ and $\varsigma_t=(x_t,y_t,0)$ such that $\bar{\varsigma}=(0,0,0)$. In practice, the real bistatic system is replaced by another bistatic system having the same \ac{PCA} but conveniently chosen to be a more accurate representation according to the considered range. For instance, if the \ac{Tx} is positioned on the array midpoint, $y_a=y_t=0$ and $x_a=-x_t$ would be the distance covered by the \ac{AUV} while the impulse reaches the target range.

In a real scenario, the whole illuminated range is observed. Hence, the range is practically split in subranges where the above approximation is adopted. The splitting in subranges must be anyway performed as the \ac{AUV} ping-to-ping motions also change along range.

\section{Simulation Results}
\label{sec:sim}

In order to assess the performances of the proposed approach, we here present results from a simulated scenario. An experiment performed in a controlled environment is presented in Section~\ref{sec:exp}.
The simulation focuses on highlighting the generality, the accuracy and the convergence of the proposed estimation procedure with no side information provided. No restrictive assumptions are made on surge and yaw.

\begin{figure}[t]
	\centering%
	\includegraphics[scale=.75,trim=0 10pt 0 10pt]{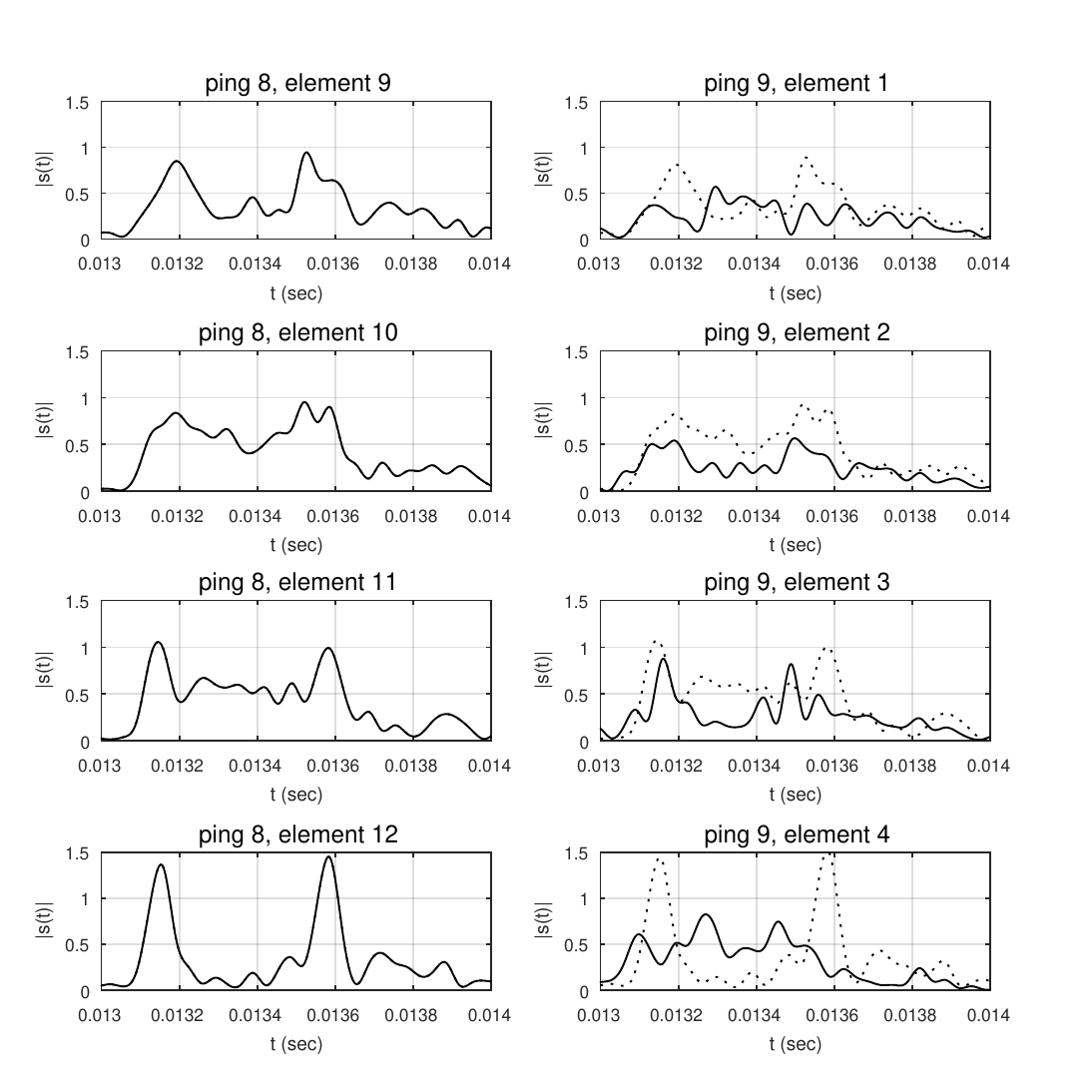}%
	\caption{%
		Raw data referred to two subsequent pings of $16$ tracks sharing $8$ overlapped phase centers being $9-16$ and $1-8$ from ping $8$ and $9$ respectively. Motion errors  $(x^{(8,9)},y^{(8,9)},\vartheta^{(8,9)})=(-3.45,2.17,4.1)\times{}10^{-3}$ are applied. The raw data with no motion errors are plotted in dotted lines. As a consequence of the surge being larger than the $\unit[2.5]{cm}$ cross-range sampling step, the correspondence between overlapping phase centers is changed. Both surge and yaw make the tracks lose their expected correlation.
	}%
	\label{fig:raw}
\end{figure}

\begin{figure}[t]
	\centering%
	\includegraphics[scale=.75,trim=0 10pt 0 10pt]{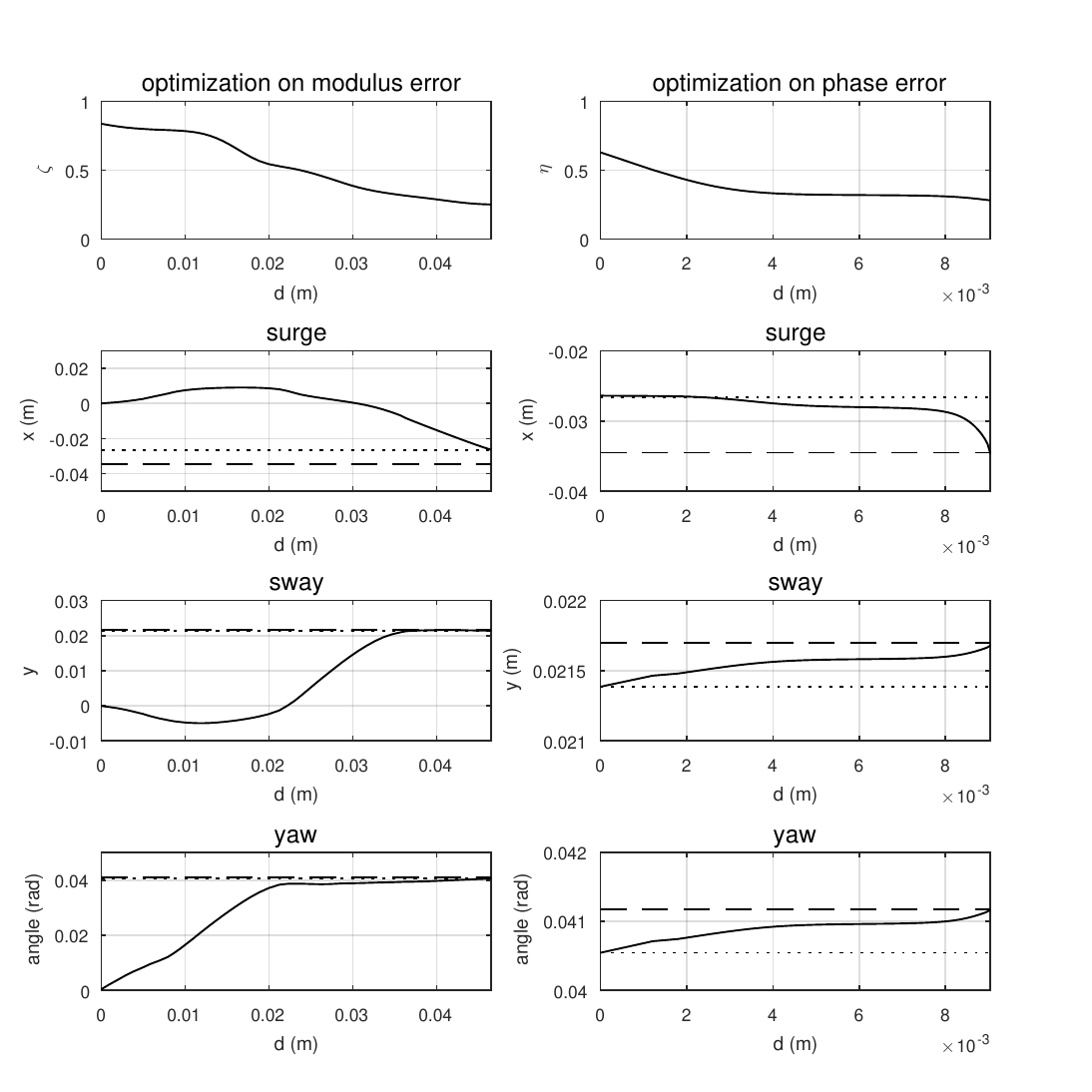}%
	\caption{%
		Error functions $\zeta$ (left) and $\eta$ (right) together with the vector components along surge, sway and yaw along the steepest descent trajectory. The optimization on $\zeta$ compensates for yaw, sway and surge in the given order. The outcome of the optimization on $\zeta$ and $\eta$ are depicted in dotted and dashed lines respectively. For visualization purposes the gradient descent trajectory has been sampled uniformly rather than according to its non-uniform sampling by means of the gradient algorithm.
	}%
	\label{fig:opt}
\end{figure}

\subsection{Interval of Convergence}

A \ac{SAS} system consisting of $16$ elements is considered, both the \ac{Tx} and the \ac{Rx} elements have $D=\unit[5]{cm}$ aperture, hence the whole array is $\unit[80]{cm}$ wide, whereas the phase centers are distributed uniformly along a $\unit[40]{cm}$ interval. As the number of the expected ping-to-ping superimposed phase centers, we considered $K=4$, so that each ping is $\unit[20]{cm}$ apart from its adjacent ones. With this setup, an \ac{AUV} moving at $\unit[2]{m/sec}$ would be capable of covering a $\unit[75]{m}$ range. The \ac{Tx} is physically located in the array midpoint, whereas it is virtually located $\unit[10]{cm}$ apart from the array midpoint position as an effect of the vehicle motion. The range observed with this assumption would be around half the maximum available range, i.e. $\unit[37.5]{m}$. Nevertheless, being a simulation with the start-and-stop approximation, we deliberately considered $\unit[10]{m}$ as midrange in order to have  a configuration where the \ac{PCA} would be less accurate, than the actual model \eqref{eq:epcar}-\eqref{eq:epcat}. The system is operating at $\unit[300]{kHz}$, while the bandwidth is $\unit[30]{kHz}$, thus resulting in $\unit[2.5]{cm}$ range resolution. The cross-range resolution is approximately equal to $D/2$, i.e. also $\unit[2.5]{cm}$. 


An instance of the objective functions $\zeta$ and $\eta$ has been plotted in Fig.~\ref{fig:srf} with respect to surge, sway and yaw separately. In more detail, the surge has been varied in the interval $[-0.2,0.2]$, i.e. the array has been moved from having its phase centers totally superimposed to the ones of the previous ping to not having any superimposed. On the other axis, a range of hypothetical motion errors has been considered. The surfaces on top qualitatively show how both the error function $\zeta$ and $\eta$ have minima on the plane diagonal, meaning that the surge can be correctly estimated. Also, $\zeta$ evaluated along the $x$ axis appears mostly convex, such that the minimum can be found by starting from $x=0$. The error function $\eta$ is steepest around the minima, but features a more remarkable non-convex behavior which makes it suitable for optimization only if a good starting point is given. Similar considerations can be done for the sway, which has been varied also in the remarkably large interval $[-0.1,0.1]$, and for the yaw. Their $\eta$ functions are not showed as they become very steep around the minima and cannot be properly visualized. 
Those results are partial as surge, sway and yaw have been considered separately, i.e., it is not showed whether those functions are convex and having the right minima when all the three motions are considered at the same time.

\begin{figure}[t]
	\centering%
	\includegraphics[scale=.75,trim=0 10pt 0 10pt]{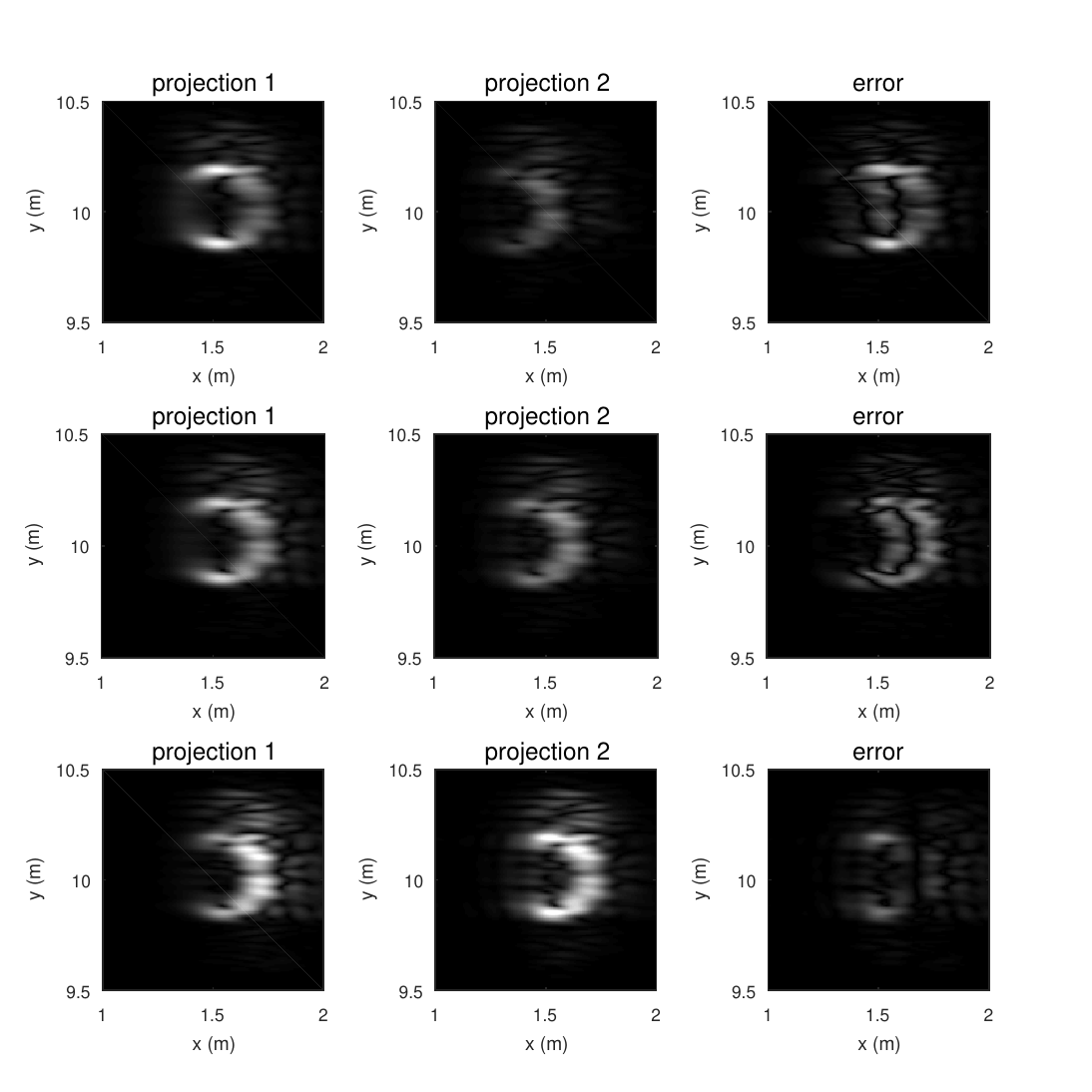}%
	\caption{%
		Projection of two successive pings on their intersection space determined as progressive approximations of their mutual displacement, being $\nicefrac{1}{2}$ (top), $\nicefrac{3}{4}$ (middle) and $1$ (bottom) of $(x^{(q,p)},y^{(q,p)},\vartheta^{(q,p)})$. Because of the approximations in computation of the projections the error is not null when the displacement is matched. Smaller details at the wavelength level responsible for the fine estimation cannot be visualize.
	}%
	\label{fig:conv}
\end{figure}

\subsection{Ping-to-ping Accuracy}

\begin{figure}[t]
	\centering%
	\includegraphics[scale=.75,trim=0 10pt 0 10pt]{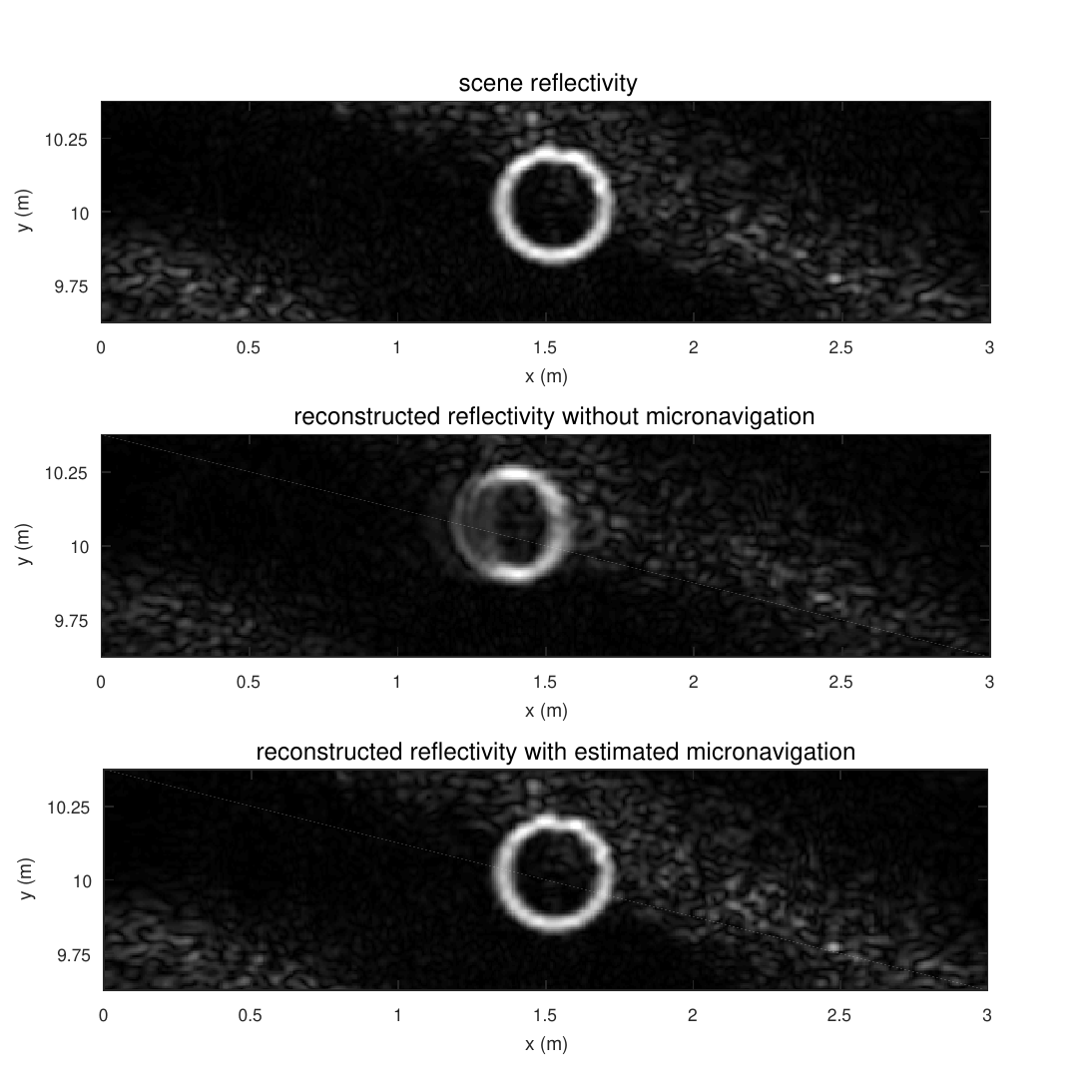}%
	\caption{%
		Absolute value of the scene complex reflectivity employed for the simulation (top), defocused image due to motion errors represented in Fig. \ref{fig:nav0} and Fig. \ref{fig:nav3} and reconstructed image by means of the proposed estimation algorithm with the errors shown in Fig. \ref{fig:nav3}. A slight increasing defocusing is observable as cross-range increases due to the cumulative errors not being zero-mean.
	}%
	\label{fig:sym}
\end{figure}

As a second experiment, we consider a single ping $q=9$ featuring motion errors with respect to its previous $p=8$, both relative to the scene depicted in Fig.~\ref{fig:sym} (top). It features a noisy background emulating sand ripples plus a phantom circular object whose diameter is $\unit[40]{cm}$. Pings $8$ and $9$ happen in the middle of the considered cross-range interval, thus facing the circular object. The motion errors are set as follows: surge $x^{(q,p)}$ equal to $\unit[-3.45]{cm}$, sway $y^{(q,p)}$ equal to $\unit[2.17]{cm}$ and yaw $\vartheta^{(p,q)}$ equal to $\unit[4.10\times{}10^{-2}]{rad}$, i.e., approximately $\unit[2.35]{deg}$. Those values are random, but their magnitude is such that other approaches such \ac{DPCA} would fail. In fact, the along-track motion error is larger than along-track sampling $D/2$ and not a multiple of $D/2$. 
The rotation error is such that the effect on raw data track cannot be described by simple delays. In Fig.~\ref{fig:raw}, raw data of tracks $9-12$ of ping $8$ are compared to tracks $1-4$ of ping $9$ . 

In Fig.~\ref{fig:cuts} we plotted the error functions $\zeta$ and $\eta$ when the considered motion errors are applied one at a time, i.e., the same as in Fig.~\ref{fig:srf}. The left column shows function $\zeta$ in solid line and $\eta$ in dotted line, whereas the right column shows the opposite. Real displacements are identified by the dashed vertical lines. The figures also show the minima which are found by the optimizations. With regard to $\zeta$ it can be seen that by starting with no a priori knowledge, i.e. $\sigma=(0,0,0)$, the optimization is convex. For surge ans sway, the size of the convexity interval is proportional to the along-track sampling step, whereas for yaw it is also inversely proportional to the number of superimposed phase centers $K$. From the figure referred to sway (middle-left) the oscillatory nature of function $\eta$ can be observed. Its period is roughly proportional to half the wavelength making it essential that the optimization on $\zeta$ has an accuracy being higher than this. 
The minima found on $\zeta$ exhibit some deviations with respect to the real displacements, which are then corrected by the optimization on $\eta$ function. The minima of $\eta$ are not equal to $0$ as the algorithms for computing the orthogonal projections and the intersection have been truncate to $5$ iterations for the sake of implementability.


From a qualitative point of view, it is interesting to have an understanding about what the projections on the intersection space represent. In Fig.~\ref{fig:conv} the evolution of the projections and their difference along the optimization process are shown. In the initial situation (top) obtained for $(x^{(q,p)},y^{(q,p)},\vartheta^{(q,p)})\times{}1/2$, the projections feature differences in shape, position and amplitude. The intermediate situation (middle) obtained for $(x^{(q,p)},y^{(q,p)},\vartheta^{(q,p)})\times{}3/4$, exhibits higher correlation between the projections but their displacements still causes the error to have the same degree of magnitude as the projections. Finally, the situation at convergence still shows some residual error and some shape differences between the projections. This example highlights the estimation robustness versus punctual differences in reflectivity observed from the two pings, as the the error functions takes into account global similarities. 

As a final step, we consider the whole optimization in the case the given displacements are happening at the same time. The optimization on function $\zeta$ gives surge equal to $\unit[-2.66]{cm}$, sway  equal to $\unit[2.15]{cm}$ and yaw $\vartheta^{(p,q)}$ equal to $\unit[4.08\times{}10^{-2}]{rad}$, whereas the subsequent optimization on $\eta$ gives $\unit[-3.49]{cm}$, $\unit[2.17]{cm}$ and $\unit[4.12\times{}10^{-2}]{rad}$ respectively. The optimization process is illustrated in Fig.~\ref{fig:opt} where the surge, sway and yaw component of the vector along the steepest descent are shown together with the function values. 

From a qualitative point of view, the resulting estimation error is highly compatible with \ac{SAS} operations, which usually require the error on sway lower than $\nicefrac{\lambda}{8}$. 
From a quantitative point of view, the assessment of the performance is quite hard as it is difficult to separate the dependency on the specific scene reflectivity from the dependency on the system parameters. Hence, estimating the error variance and bias could be only approached statistically. Nevertheless, it could be argued that both theoretical and/or statistical results on simulated data might focus on aspects which could be remarkably overtaken from other issues when moving to real data. For this reason, we believe it is not worth pursuing a further statistical estimation error analysis before assessing the working principle on real data. 

\begin{figure}[t]
	\centering%
	\includegraphics[scale=.75,trim=0 10pt 0 10pt]{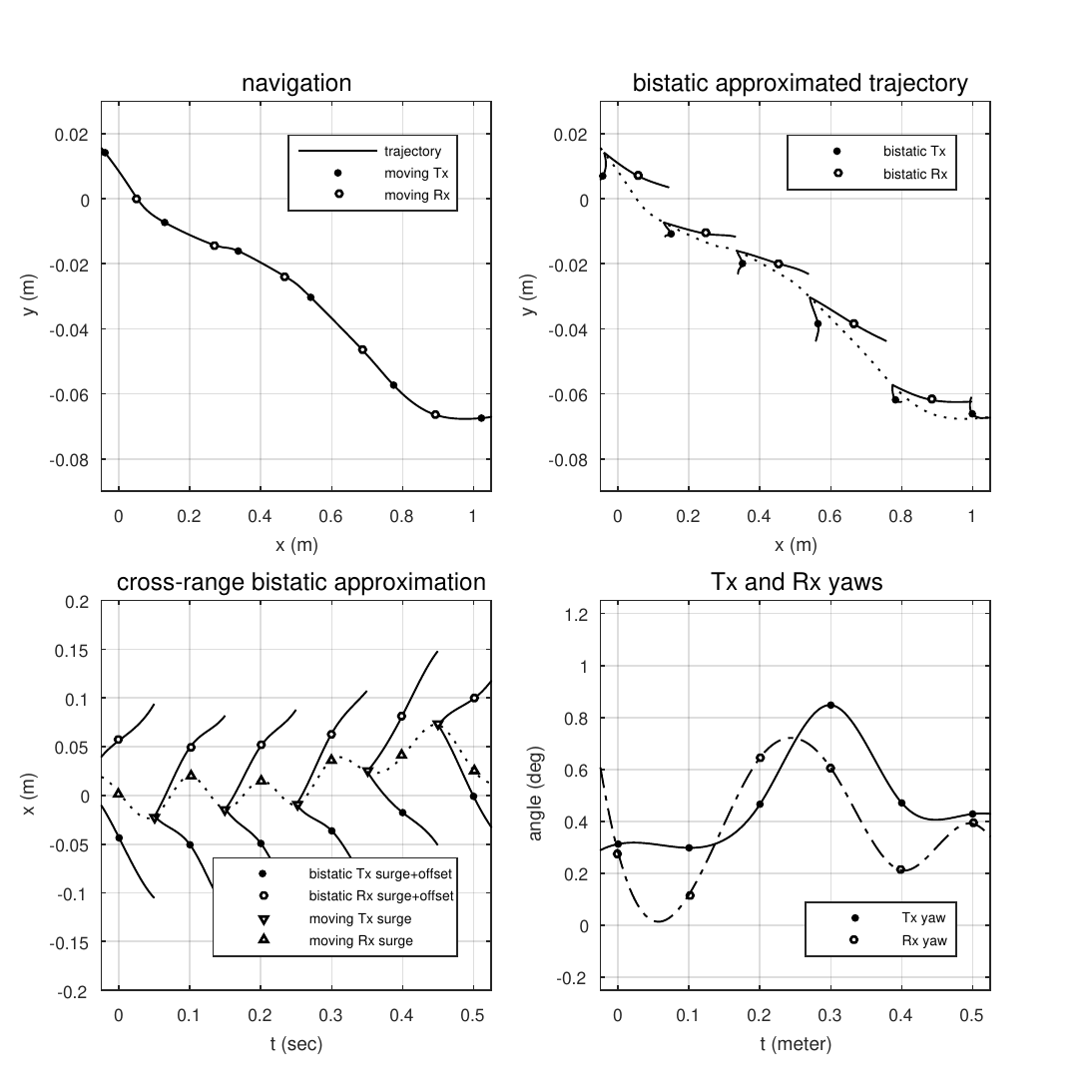}%
	\caption{%
		Navigation employed in the proposed simulation for an \ac{AUV} moving at $\unit[2]{m/s}$ approximately along the azimuth (x) axis. The trajectory on the azimuth-range plane  employed for obtaining the raw data is illustrated (the top-left), where the highlighted points implicitly specify the range of interest. The \ac{Rx} position refers to the center of the array. The replacement bistatic \ac{Tx}/\acp{Rx} pairs employed for the reconstruction are also shown (top-right) together with the lines representing their positions for different ranges. The deviation from the prescribed track along the azimuth axis is represented in the bottom-left plot. The bistatic \ac{Tx} and \ac{Rx} are $\unit[10]{cm}$ apart as it was along the prescribed track. Yaw for both the \ac{Tx} and \ac{Rx} is also shown (bottom-right).
	}%
	\label{fig:nav0}
\end{figure}

\begin{figure}[t]
	\centering%
	\includegraphics[scale=.75,trim=0 10pt 0 10pt]{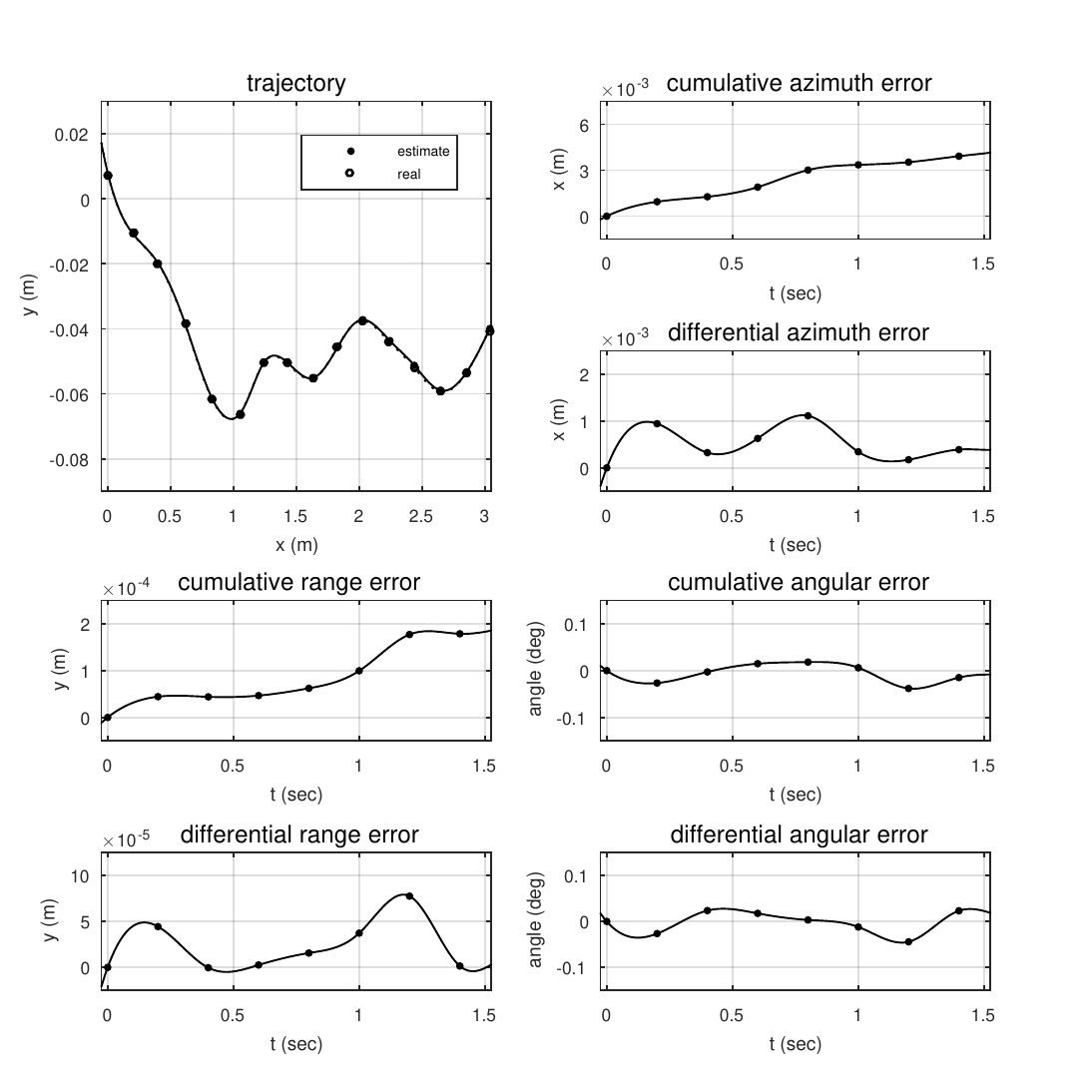}%
	\caption{%
		Full trajectory employed in the simulation and its estimate (top-left). Differential and cumulative errors with respect to the the along-path, range and angular directions. The obtained errors are compatible with \ac{SAS} operations and allow for an accurate reconstruction of the scene reflectivity although no prior has been exploited in order to determine the trajectory. Small biases seem to occur for surge and sway but the simulation is not statistically relevant.
	}%
	\label{fig:nav3}
\end{figure}

\subsection{Trajectory Accuracy}

As a final simulation, we provide the estimated trajectory along a $\unit[3]{m}$ path covered by $16$ pings. The considered scene reflectivity is again the one depicted in the top of Fig.~\ref{fig:sym}. The reconstructed scene with no navigation is plotted in the middle figure. 
To make the simulation more realistic, the generation of raw data takes into account the different location and rotation of the \ac{Tx} with respect to the \ac{Rx}, whereas the reconstruction is performed by means of the proposed bistatic model \eqref{eq:epcar}-\eqref{eq:epcat} assuming that the expected \ac{Tx} to \ac{Rx} distance along the cross-range direction is known and equal to $\unit[10]{cm}$. 
The considered random trajectory is partially shown in Fig.~\ref{fig:nav0} (top-left), together with a graphical explanation of the employed bistatic model. In the top-right figure the position of the \ac{Tx} and array midpoint are explicitly shown.
The result of the trajectory estimation are shown in Fig.~\ref{fig:nav3}. The differential and cumulative errors on surge and sway might suggest small biases, which can be neglected as the final sway error is still only about $\nicefrac{\lambda}{25}$. Finally, the obtained trajectory is employed to reconstruct the scene in the top figure of Fig.~\ref{fig:sym}. The obtained image is qualitatively comparable to the original image and the whole estimation procedure fulfils the initial goal of estimating motion errors with no priors.

\section{Experimental Results}
\label{sec:exp}

Given the peculiarity of the proposed estimation technique and its complex algebraic structure, assessing its validity on real data is a major concern. An experiment in a controlled environment has been designed at this aim. Nevertheless, the a priori knowledge of the sensor locations could not be provided by means of the available equipment. For this reason, a specific manner to validate the experiment has been identified and designed.


\subsection{System Setup}

\begin{figure}[t]
	\centering%
	\includegraphics[scale=1.068,trim=0 10pt 0 10pt]{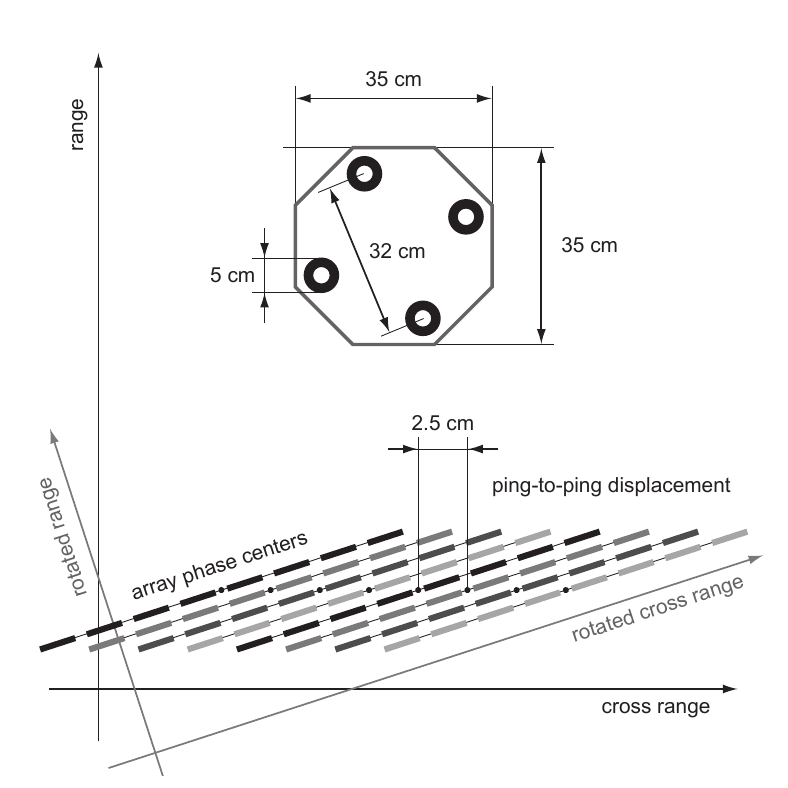}%
	\caption{%
		Representation of the scene employed for the designed experiment. The array midpoint is moved along the cross-range axis. However, as the rotation of the array along the whole trajectory is not known, the trajectory obtained by summing all the differential displacements is aligned with the first ping. As a consequence, the estimated trajectory may result rotated with respect to the cross-range axis.
	}%
	\label{fig:scn}
\end{figure}

For this purpose, a water tank being $\unit[3]{m}$ large, $\unit[4]{m}$ long and $\unit[2]{m}$ deep was available. In order to avoid reflections from the concrete wall and bottom, a target consisting of $4$ metal reflectors lying on an octagonal plastic surface parallel to the surface and mounted on a wooden support at approximately $\unit[1]{m}$ from the bottom has been considered. The geometry of the object is detailed in Fig.~\ref{fig:scn}. A \ac{SAS} system consisting of $8$ \acp{Rx} of $\unit[5]{cm}$ aperture and spaced by $\unit[5]{cm}$ and a $\unit[5]{cm}$ aperture \ac{Tx} positioned at the midpoint of the array has been provided by Hydrason Solutions. The \ac{SAS} system is capable of operating between $\unit[20]{kHz}$ and $\unit[170]{kHz}$. The tank is provided with an industrial plotter which has been employed in order to move the \ac{SAS} system along a straight line and perform the acquisitions by the start-and-stop approach. 
According to the specifications, the plotter provide a submillimetre accuracy. However, as the plotter engine produces a non-negligible noise, it had to be switched on and off throughout the whole acquisition process. This operation has caused the displacement accuracy to be inaccurate. 
Other causes of inaccuracy might have come from other mechanical non-idealities, such as submillimetre deviations in the supporting guide.

The \ac{SAS} system has been attached to the plotter by means of a pole and mounted on a rotating plate capable of providing roll and yaw. Roll has been adjusted in order to make the beam illuminate the object before touching the tank bottom, whereas yaw has been adjusted to be null. However, these setups have been manually arranged, so no apriori accurate information can be assumed for those parameters. Hence, the positioning of the array throughout all its displacement along the guide can feature both yaw and roll. Finally, the relatively closeness of the target object makes the \ac{PCA} assumption weak and the reflectivity to vary along the trajectory, thus causing a lack of coherence from ping to ping.

The \ac{SAS} raw data have been collected by moving the system by steps equal to the spacing between phase centers, i.e. $\unit[2.5]{cm}$, being also the expected along-track resolution, i.e. $D/2$. 
An interval of $\unit[80]{cm}$ plus the length of the array \acp{PCA}, i.e. $\unit[20]{cm}$, has been covered by $32$ pings, thus getting a synthetic aperture of $\unit[1]{m}$. As far as the physical parameters are concerned, a $\unit[30]{kHz}$ chirp of duration equal to $\unit[3]{msec}$, central frequency equal to $\unit[105]{kHz}$ and sampled at $\unit[1]{MHz}$ has been employed. After pulse compression, demodulation and filtering, a downsampling by $33$ has been performed, thus getting a cross-track resolution equal to $\unit[2.45]{cm}$.


\subsection{Validation Methodology}

\begin{figure}[t]
	\centering%
	\includegraphics[scale=1.068,trim=0 10pt 0 10pt]{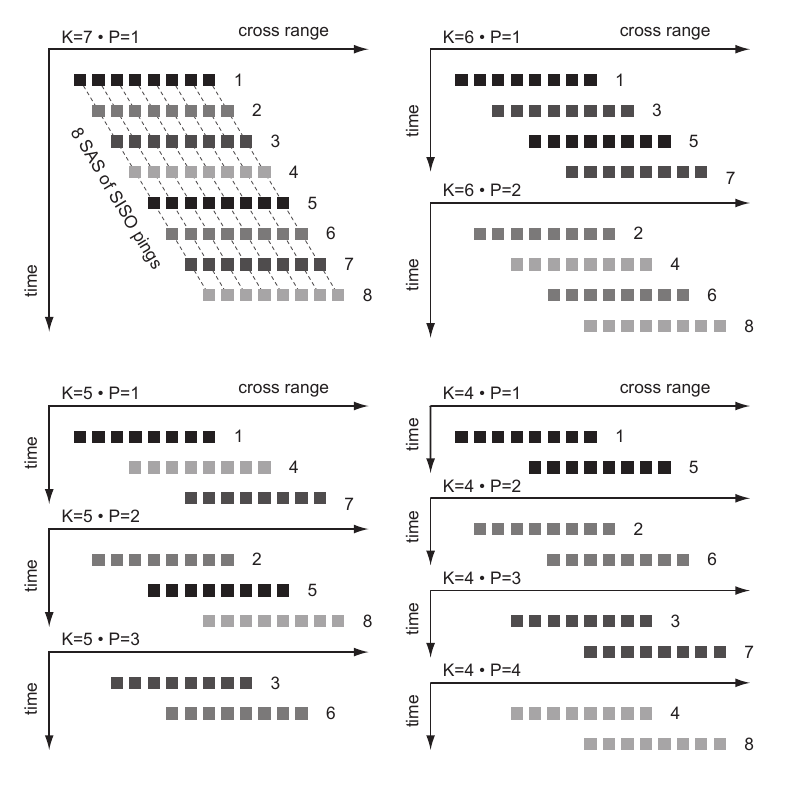}%
	\caption{%
		Schematic representation of the different arrangements of the $32$ acquired pingsm exemplified on the first $8$ ones. By varying the number of overlapping phase centers $K$, $N-K$ \ac{SAS} systems can be obtained, each generating a subsampled trajectory estimation. By combining those, $K$ fully sampled trajectories are estimated. The pings can be also organized in $8$ \ac{SAS} of \ac{SISO} systems.
	}%
	\label{fig:cmb}
\end{figure}

By means of $M$ pings with $K=N-1$, $MN$ different observations are collected. So, for this experiment $256$ tracks are acquired, which can be used in various ways. For instance, given an array whose \acp{PCA} covers uniformly $\unit[20]{cm}$, the considered synthetic aperture of $\unit[1]{m}$ can be equivalently covered with $5$ pings and $K=0$ or $10$ pings and $K=4$. By means of the considered $32$ pings, two main tasks can be accomplished: (i) the $32$ pings can be subsampled in order to produce $P=1,\ldots,N-K$ different motion estimates for $K=1,\ldots,7$; (ii) by selecting one of the $8$ \ac{Rx} at a time, 8 new synthetic aperture can be obtained as a collection of the corresponding \ac{SISO} systems. Both these tasks are illustrated in Fig.~\ref{fig:cmb}. The simple case $K=7$ allowing for the single phase $P=1$ is rearranged in order to obtain the case $K=4,5,6$ with all the corresponding different phases. Also, the case $K=7$ is reorganized in virtual arrays of \ac{SISO} systems.

The benefit of the \ac{SAS} made up of \ac{SISO} systems is the following. Given that the elements are moved on a straight line, no sway can be present. A small rotation of each element is not actually impacting on the imaging task, so surge is the only motion error to be identified. Hence, the $8$ images which can be obtained the \ac{SISO} approach should be similar apart from a slight warping in the track direction. Hence they can be employed as initial ground truth. Conversely, the multiple motion estimates at different sampling steps can be exploited as follows. For each value of $K$, each of the $P=1,\ldots,N-K$ estimates represents a subsampling of the full sampled trajectory with different initial phase. Hence, they can be integrated in order to obtain a single motion estimate for each value of $K$. Then the obtained $K$ trajectories can be applied separately on each of the $8$ \ac{SAS} based on the \ac{SISO} approach, or to each one of the $8$ \ac{SAS} based on the proper  array, which can be referred to as \ac{SIMO} system. This would lead to $K\times{}8$ images for the \ac{SISO} and for \ac{SIMO} case respectively. Ideally, if the motion estimation procedure works accurately, all those reconstructed complex reflectivities should be equal.

\begin{figure}[t]
	\centering%
	\includegraphics[scale=.75,trim=0 10pt 0 10pt]{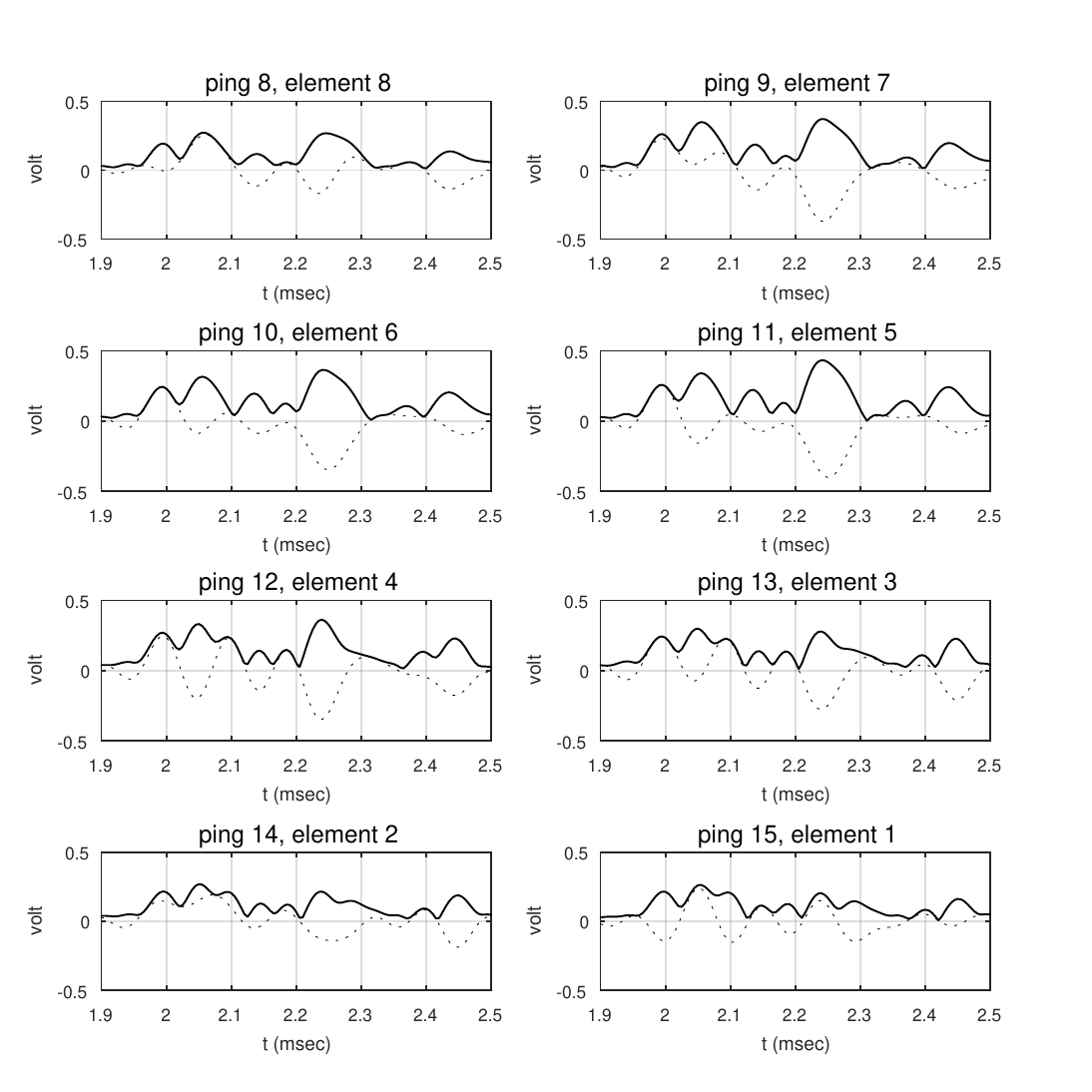}%
	\caption{%
		Raw data from the experiment corresponding to $8$ different \ac{SISO} systems which share the same phase center apart from surge and sway motions. Both the absolute value (solid-line) and the imaginary part (dotted line) are plotted. The four tracks in the bottom do not feature correlation with the four on top because of a major surge motion between ping $11$ and $12$. The variation of the imaginary part across all the tracks is due to sway.
	}%
	\label{fig:png}
\end{figure}

\subsection{Motion Errors}

With regard to surge, it was known that in correspondence of ping $12$ a major misalignment occurred. In Fig.~\ref{fig:png} all $8$ tracks corresponding to a single location have been plotted. Tracks $8-11$ and $12-15$ are not correlated as a consequence of this surge event. Moreover, a finer comparison of the tracks would reveal that other differences are present even within each group. Those can be attributed to a lack of calibration among the array elements and also to the fact that the 8 \ac{SISO} systems are supposed to share the same phase center but practically look at the scene from different angles. 

When considering the \ac{SIMO} systems, surge is the same as for the \ac{SISO} ones, whereas sway can appear as an effect of the initial orientation of the array. This is explained in Fig.~\ref{fig:scn}, where it is shown that the midpoint of a rotated array moves diagonally with respect to the reference system aligned with the array. Hence, the angle of this diagonal line is a measurement for the array orientation with respect to the scene reference system.


\begin{figure}[t]
	\centering%
	\includegraphics[scale=.75,trim=0 10pt 0 10pt]{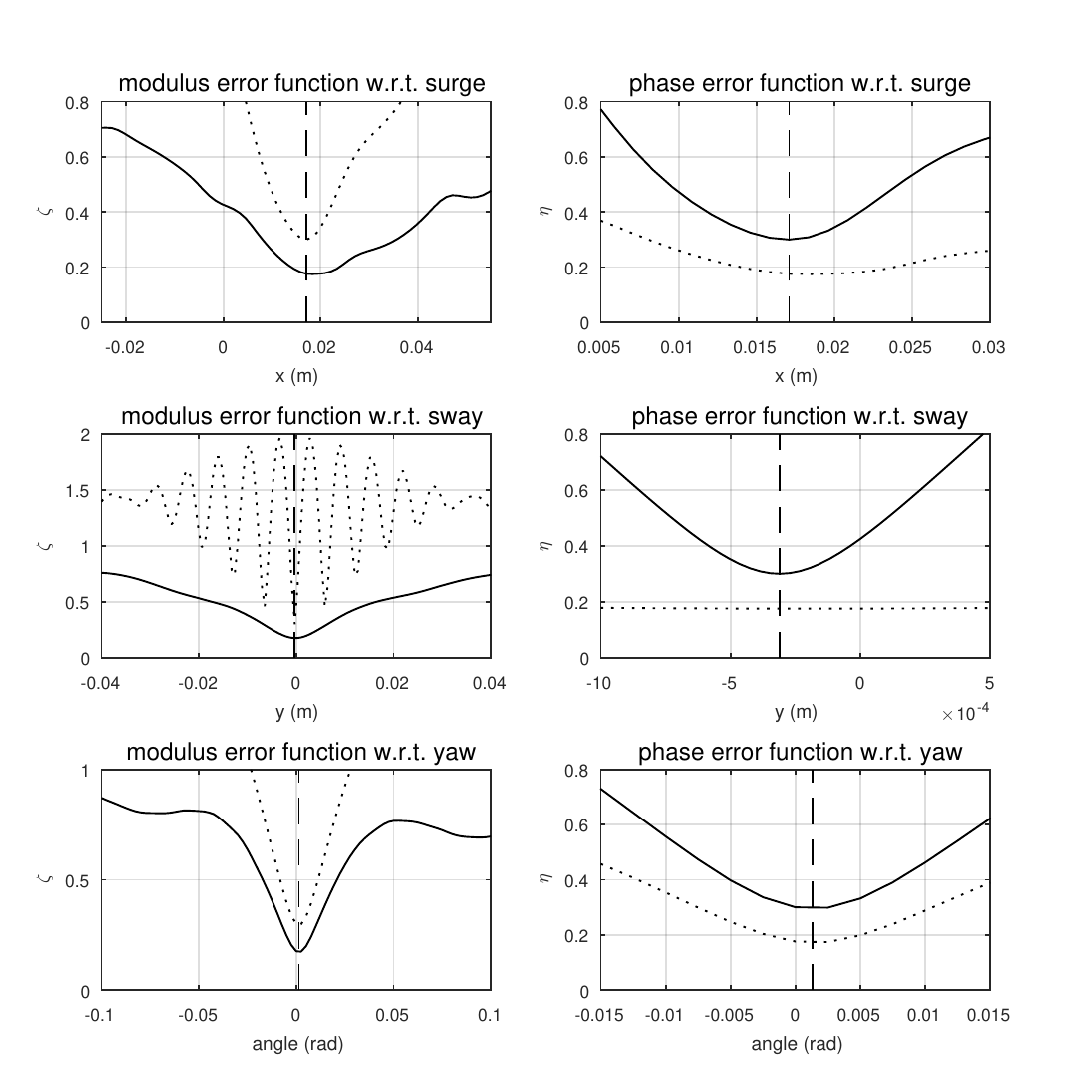}%
	\caption{%
		Error functions $\zeta$ (solid line on the left and dotted line on the right) and $\eta$ (dotted line on the left and solid line on the right) on real data exemplified on surge, sway and yaw separately (top, middle and bottom respectively). Dashed vertical lines identify the estimated displacement. The shape of the functions and their convexity matches the error functions simulated in Fig.~\ref{fig:cuts}.
	}%
	\label{fig:crv}
\end{figure}

\subsection{Estimated Trajectory}

Before performing the actual trajectory estimation by means of the optimizations, it is worth analysing the error functions in order to assess that they feature the same characteristics as in Fig.~\ref{fig:cuts}. Error function $\zeta$ and $\eta$ with respect to pings $11$ and $12$ are plotted in Fig.~\ref{fig:crv}. All the macroscopic features highlighted for emulated data are replicated for real data. The functions are convex on comparable intervals and $\eta$ with respect to sway oscillates with period equal to half the wavelength (the oscillation is much more visible here as the central frequency is $\unit[105]{kHz}$ rather than $\unit[300]{kHz}$). For the considered ping, surge, sway and yaw are equal to $\unit[1.71]{cm}$, $\unit[-3.1]{mm}$ and $\unit[1.35\times{}10^{-3}]{rad}$ respectively. The yaw motion error is less than $\nicefrac{1}{10}$ of one degree, so it can be considered negligible.

\begin{figure}[t]
	\centering%
	\includegraphics[scale=.75,trim=0 10pt 0 10pt]{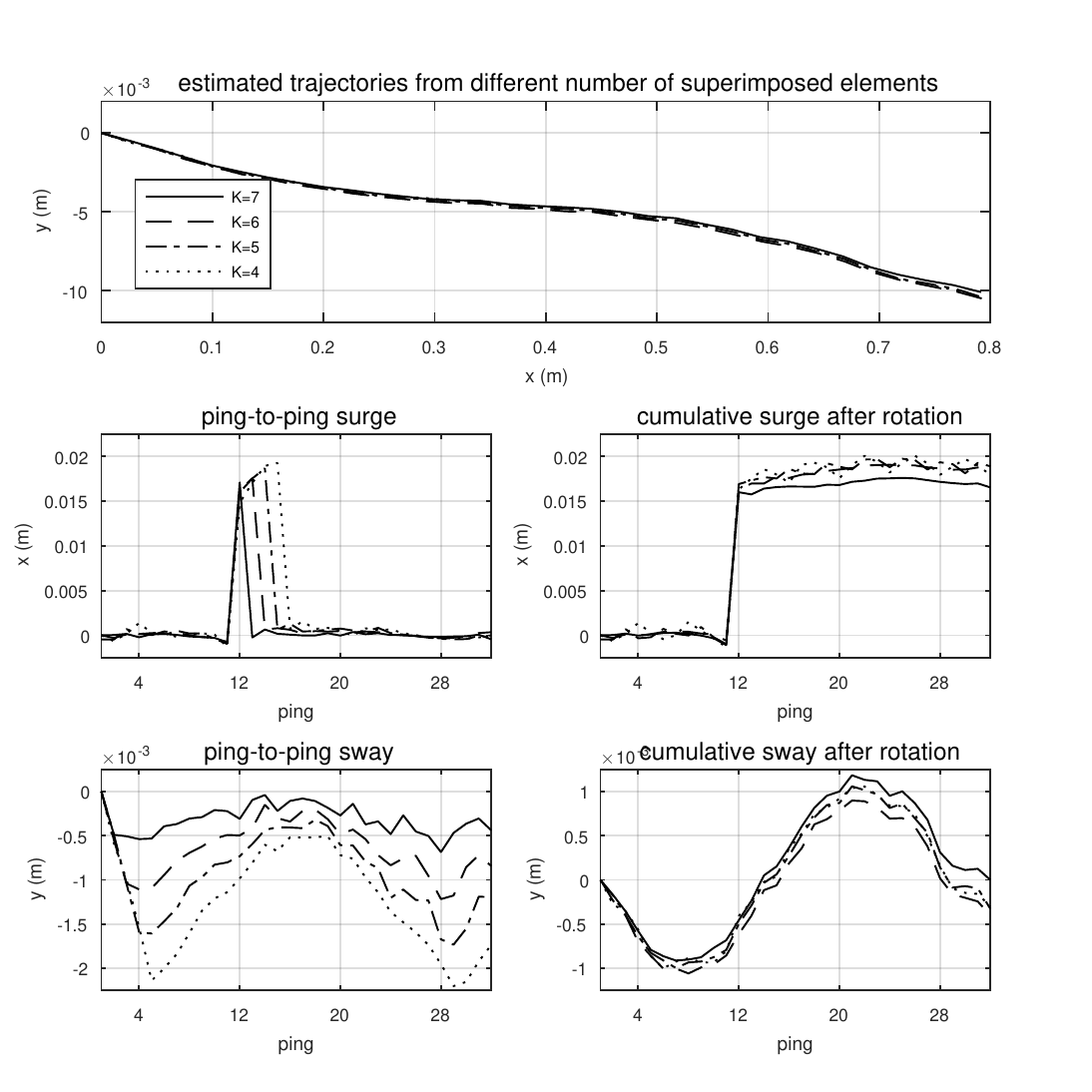}%
	\caption{%
		Estimated trajectories (top) from the four interlaced differential displacements (bottom left) obtained by means of the ping arrangements represented in Fig.~\ref{fig:cmb}. The shown cumulative displacement along the track (bottom right) have been rotated by the angle determined line approximating the trajectory.
	}%
	\label{fig:trj}
\end{figure}

Moving from one ping to all the tracks, the results are shown in Fig.~\ref{fig:trj}. By properly combining the differential displacements obtained for all the different phases illustrated in Fig.~\ref{fig:cmb}, a trajectory estimation can be obtained for each considered value of $K$, thus resulting in $4$ trajectories plotted in Fig.~\ref{fig:trj} (top). The differential surge and sway are also shown (left-bottom). It is worth noting that all the phases different from $1$ have an undetermined starting point which has been numerically tuned to maximize the match with the trajectory obtained for $K=1$. Also, the differential motion errors represented in the figure are interlaced. As the trajectory lays  along a diagonal path with respect to the track, the cumulative displacements have been rotated by an angle equal to the opposite of the diagonal orientation, i.e. $\unit[1.27\times{}10^{-2}]{rad}$ or $\unit[0.73]{deg}$, and the same angle has been added to the yaw. The resulting cumulative displacements (right-bottom) give a hint about how the accuracy might decrease when the number of overlapping phase centers decreases. The overall deviation is enclosed in less than $\unit[5]{mm}$ and $\unit[0.5]{mm}$ for surge and sway respectively. 


\subsection{Reconstructed Reflectivity}

As mentioned before in this Section, the absolute value of the reconstructed complex reflectivities obtained from the \ac{SISO} wise \ac{SAS} have to be equal apart slight deformation due to surge. Absolute values have been used as an initial way to assess the correctness of the estimation approach. Nevertheless, it is more proper to look at the complex reflectivities and verify that the both the real and the imaginary parts, 
are equal for both the \ac{SISO} and the \ac{SIMO} \ac{SAS} for all the $8$ set of pings having $K=0$.

\begin{figure}[t]
	\centering%
	\includegraphics[scale=.75,trim=0 10pt 0 10pt]{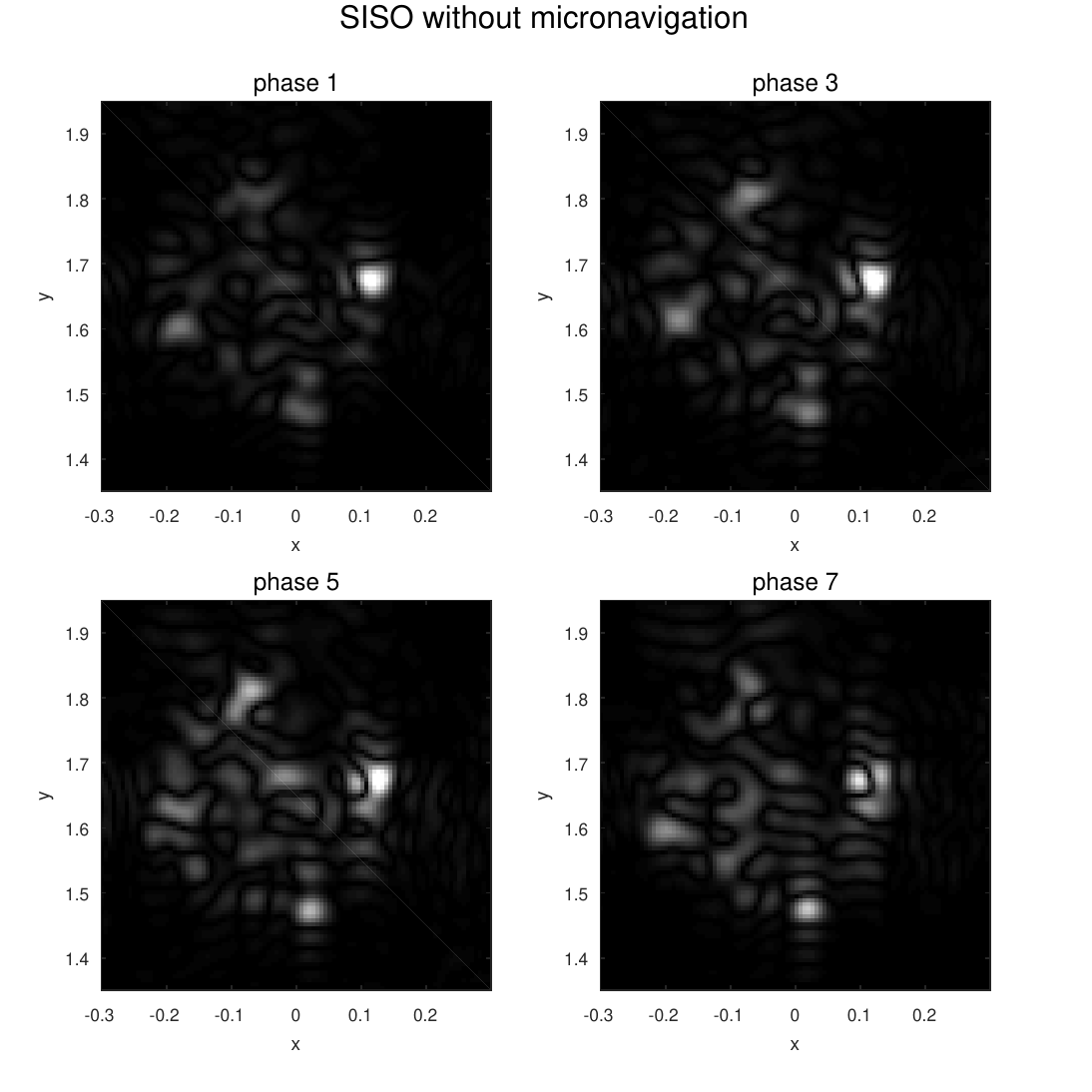}%
	\caption{%
		Absolute value of the real part of the complex reflectivity reconstructed from the four different independent \ac{SAS} systems constituted by a collection of \ac{SISO} bistatic systems. Those systems are identified by considering a selection of the $8$ group of pings gathered in diagonal in the top-left plot in Fig.~\ref{fig:cmb}. No motion compensation has been applied.
	}%
	\label{fig:siso_nonav}
\end{figure}

\begin{figure}[t]
	\centering%
	\includegraphics[scale=.75,trim=0 10pt 0 10pt]{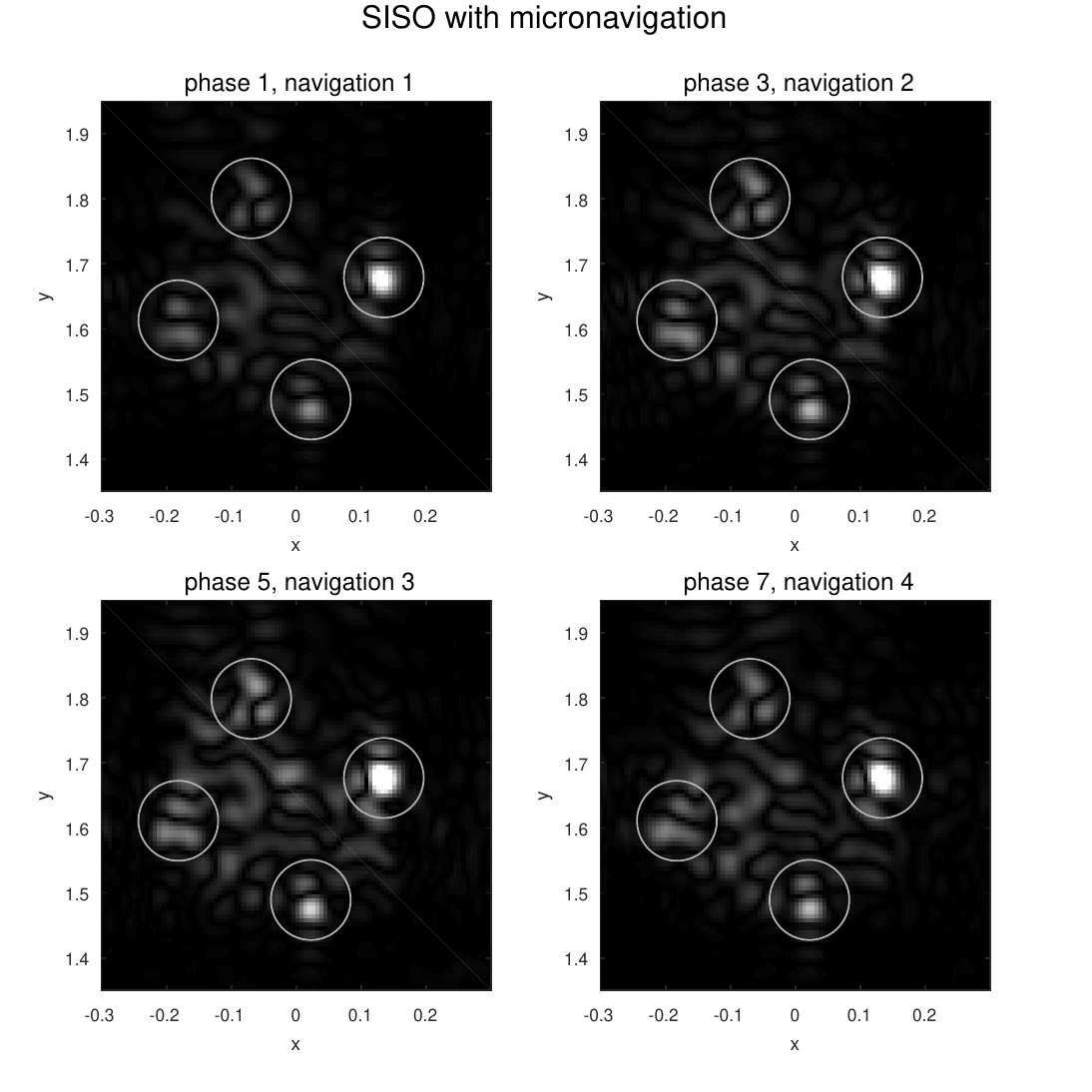}%
	\caption{%
		Absolute value of the real part of the complex reflectivity relative to the cases shown in Fig.~\ref{fig:siso_nonav} after motion compensation according to the trajectories estimated through the four ping arrangement illustrated in Fig.~\ref{fig:cmb}. The highlighted areas correspond to the reflectors illustrated in Fig.~\ref{fig:scn}, which exhibit mutual consistency meaning that the motion estimation procedure has been performed correctly.
	}%
	\label{fig:siso_nav}
\end{figure}

Following this, in Fig.~\ref{fig:siso_nonav} absolute value of the real part of the reconstructed reflectivities for the \ac{SISO} \ac{SAS} have been plotted for $4$ of the $8$ possible set of pings with no motion compensation. Despite the fact the their absolute values are fairly similar, the phase is different as the \ac{SISO} element phase centers are located on different parallel lines. In Fig.~\ref{fig:siso_nonav} the same plots are proposed with respect to the \ac{SIMO} \ac{SAS}. The motion compensated version of these cases have been represented in Fig.~\ref{fig:siso_nav} and Fig.~\ref{fig:simo_nav} respectively, each by means of a different estimated trajectory in order to show that the error among them is negligible. The areas where the coherent metallic reflectors are located have been highlighted to emphasize the consistency across the various cases. As expected, the rest of the area features some differences for the intrinsic lack of coherence due to the close range.

\section{Conclusion}
\label{sec:concl}

A novel motion compensation technique for \ac{SAS} capable of identifying surge, sway and yaw with no restrictions and no prior information from inertial sensors has been introduced. The proposed approach is based on the comparison between the projections on the intersection subspace of different observations. The effectiveness of the technique has been proved by extensive analysis and simulations together with an experiment on real data.

\section{Acknowledgements}

This work was supported by the Engineering and Physical Sciences Research Council (EPSRC)
Grant number EP/J015180/1 and the MOD University Defence Research Collaboration
in Signal Processing.



\newpage

\begin{figure}[t]
	\centering%
	\includegraphics[scale=.75,trim=0 10pt 0 10pt]{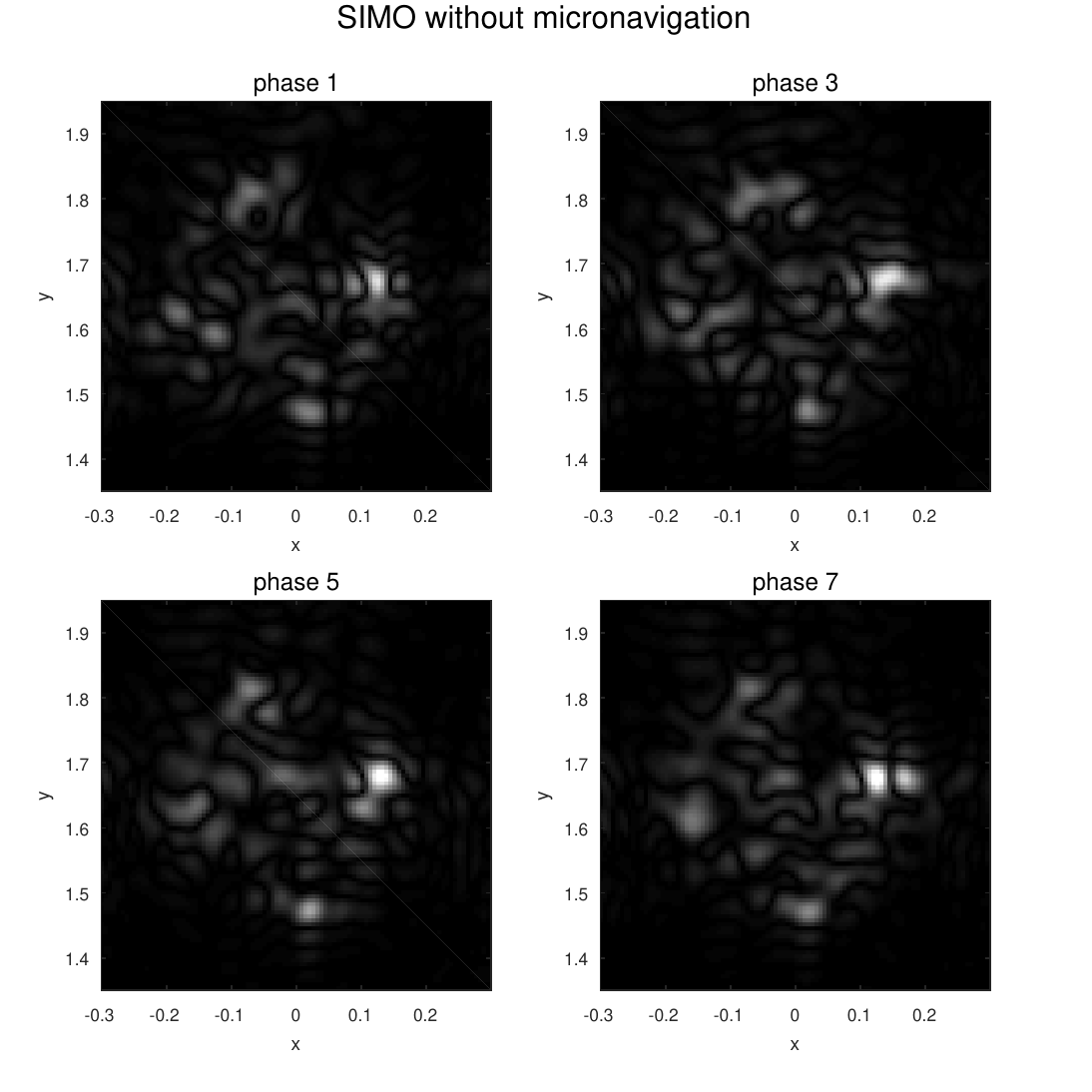}%
	\caption{%
		Absolute value of the real part of the complex reflectivity reconstructed from the four different independent \ac{SAS} systems constituted by a collection of \ac{SIMO}  systems, being pings $M+kN$, $k=0,1,\ldots$ and $M=1,3,5,7$, each having the number of overlapping phase centers $K$ equal to $0$. No motion compensation has been applied.
	}%
	\label{fig:simo_nonav}
\end{figure}

\begin{figure}[t]
	\centering%
	\includegraphics[scale=.75,trim=0 10pt 0 10pt]{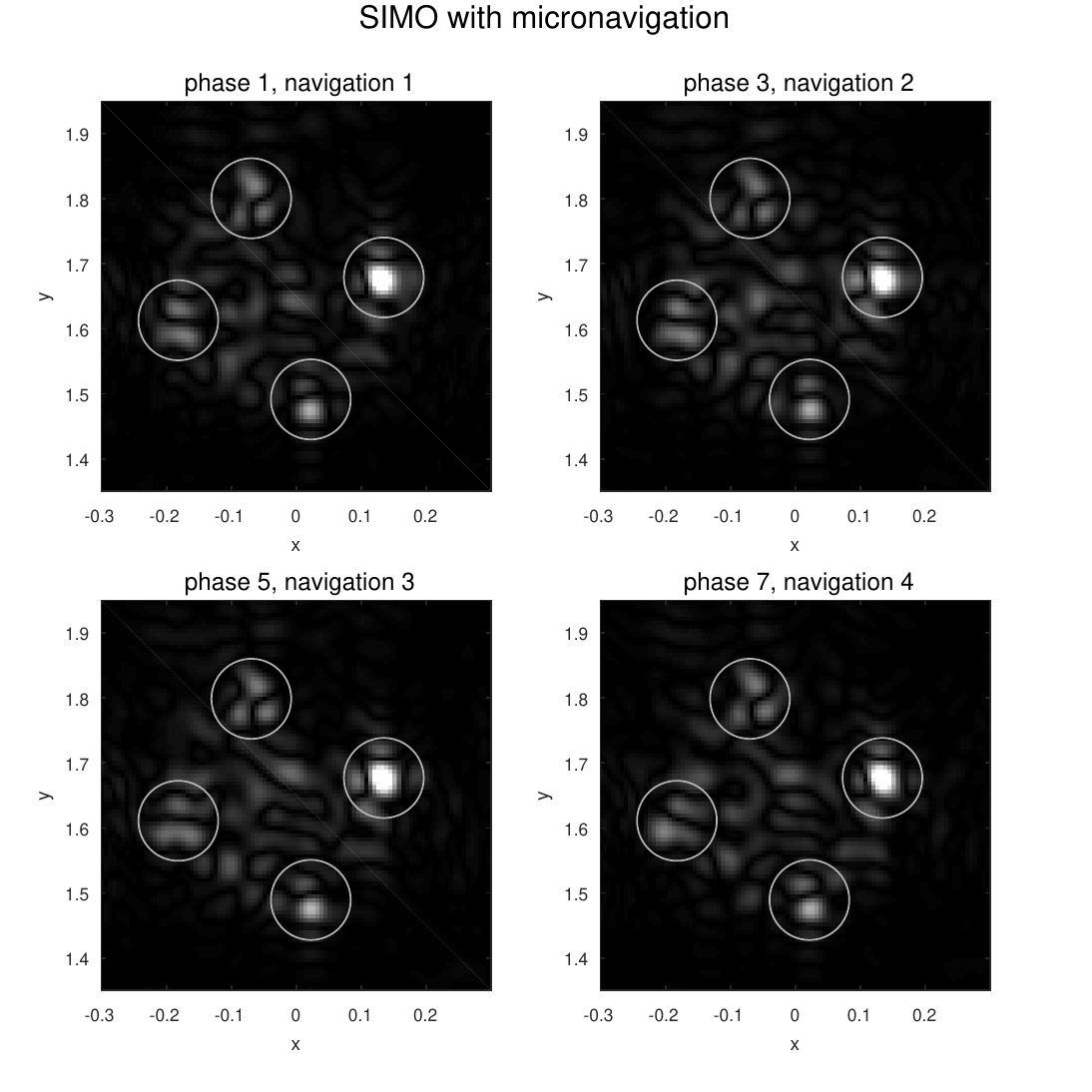}%
	\caption{%
		Absolute value of the real part of the complex reflectivity relative to the cases shown in Fig.~\ref{fig:simo_nonav} after motion compensation according to the trajectories estimated through the four ping arrangement illustrated in Fig.~\ref{fig:cmb}. The highlighted areas corresponding to the reflectors illustrated in Fig.~\ref{fig:scn} have mutual consistency and are also consistent with respect to the cases shown in Fig.~\ref{fig:siso_nav}.
	}%
	\label{fig:simo_nav}
\end{figure}


%
%
%


\end{document}